\documentclass[fleqn,usenatbib]{mnras}

\usepackage{newtxtext,newtxmath}

\usepackage[T1]{fontenc}

\DeclareRobustCommand{\VAN}[3]{#2}
\let\VANthebibliography\thebibliography
\def\thebibliography{\DeclareRobustCommand{\VAN}[3]{##3}\VANthebibliography}

\usepackage{amsmath}
\usepackage{graphicx}
\usepackage{hyperref}
\usepackage[position=top]{subfig}
\usepackage[parfill]{parskip}
\usepackage{array}
\usepackage{booktabs}
\usepackage[dvipsnames]{xcolor}
\usepackage{tikz}
\usepackage{todonotes}
\usepackage{caption,tabularx,booktabs}
\usepackage{mwe}
\usepackage{booktabs,multirow}
\usepackage{float}
\restylefloat{table}

\newlength{\tempheight}
\newlength{\tempwidth}

\newcommand{\rowname}[1]
{\rotatebox{90}{\makebox[0.1\tempheight][c]{\textbf{#1}}}}

\newcommand{\columnname}[1]
{\makebox[\tempwidth][c]{\textbf{#1}}}

\title[Stellar ex-situ mass fraction inference from observables]{ERGO-ML: Towards a robust machine learning model for inferring the fraction of accreted stars in galaxies from integral-field spectroscopic maps
}

\author[E. Angeloudi et al.]{Eirini Angeloudi,$^{1, 2}$\thanks{E-mail: eirini@iac.es}
Jes\'us Falc\'on-Barroso,$^{1, 2}$ Marc Huertas-Company$^{1, 2, 3, 4}$, Regina Sarmiento$^{1, 2}$,\newauthor Annalisa Pillepich$^{5}$, Daniel Walo-Mart\'in$^{1, 2}$ and Lukas Eisert$^{5}$
\\
$^{1}$Instituto de Astrof\'isica de Canarias, C. V\'ia L\'actea, 1, E-38205 La Laguna, Tenerife, Spain \\
$^{2}$Universidad de la Laguna, dept. Astrof\'isica, E-38206 La Laguna, Tenerife, Spain\\
$^{3}$Universit\'e Paris-Cit\'e, LERMA - Observatoire de Paris, PSL, Paris, France\\
$^{4}$SCIPP, University of California, Santa Cruz, CA 95064, USA\\
$^{5}$Max Planck Institute for Astronomy,
Königstuhl 17, 69117 Heidelberg, Germany
}

\date{Accepted 2023 May 29. Received 2023 May 15; in original form 2023 April 03}

\pubyear{2023}

\begin{document}
\label{firstpage}
\pagerange{\pageref{firstpage}--\pageref{lastpage}}
\maketitle

\begin{abstract}

\noindent
Quantifying the contribution of mergers to the stellar mass of galaxies is key for constraining the mechanisms of galaxy assembly across cosmic time. However, the mapping between observable galaxy properties and merger histories is not trivial: cosmological galaxy simulations are the only  tools we have for calibration. We study the robustness of a simulation-based inference of the ex-situ stellar mass fraction of nearby galaxies to different observables -- integrated and spatially-resolved -- and to different galaxy formation models -- IllustrisTNG and EAGLE -- with Machine Learning. We find that at fixed simulation, the fraction of accreted stars can be inferred with very high accuracy, with an error $\sim5$ per cent (10 per cent) from 2D integral-field spectroscopic maps (integrated quantities) throughout the considered stellar mass range. A bias (> 5 per cent) and an increase in scatter by a factor of 2 are introduced when testing with a different simulation, revealing a lack of generalization to distinct galaxy-formation models. Interestingly, upon using only stellar mass and kinematics maps in the central galactic regions for training, we find that this bias is removed and the ex-situ stellar mass fraction can be recovered in both simulations with <15 per cent scatter, independently of the training set's origin. This opens up the door to a potential robust inference of the accretion histories of galaxies from existing Integral Field Unit surveys, such as MaNGA, covering a similar field of view (FOV) and containing spatially-resolved spectra for tens of thousands of nearby galaxies. \\

\end{abstract}




\section{Introduction}

Large-volume cosmological hydrodynamical galaxy simulations have proved to be an invaluable ally over the past years in acquiring a deeper understanding of the physical processes that govern galaxy formation and evolution. From helper tools struggling to reproduce fundamental properties of their observable counterparts, such simulations have evolved a long way into significantly accurate models of our current perception of the observable Universe \citep{Somerville_2015}. 

The increased agreement between simulation outputs and observational datasets is mainly reflected in their ability to reproduce, among others, the galaxy stellar mass functions, key observed galaxy scaling laws as well as realistic galaxy morphologies \citep{Vogelsberger2020}. A main contributor for this agreement is the inclusion of sub-grid physics recipes in the cosmological scheme, chiefly feedback from star formation and from super massive black holes i.e. AGN, which allow to overcome the unavoidable limitations on spatial ($\sim$ kpc) and mass resolution ($\sim 10 ^{6} \rm{M}_\odot$).
These sub-grid recipes, which together with all other physical processes solved in galaxy simulations are referred to as galaxy-formation models, allow the modeling of a number of phenomena that in reality occur below the resolution limit
and are included in all major current projects to simulate galaxies: e.g. EAGLE \citep{eagle, eagle2}, Illustris \citep{illustris1, illustris2, illustris3, illustris4}, IllustrisTNG \citep{2018MNRAS.480.5113M, 2018MNRAS.475..676S, 2018MNRAS.475..648P, 2018MNRAS.475..624N, 2018MNRAS.477.1206N} and SIMBA \citep{simba}, to name a few. \looseness-2

The improvement in realism of cosmological galaxy simulations, in addition to their capability in returning many thousands of simulated galaxies at once, has allowed the emergence of works based on machine learning (ML). A number of these, for example, have aimed at calibrating the relations between galaxy observables and physical properties \citep[e.g.][]{Huertas-Company_2018, pearson, bottrell_2021, bickley_2021, Ntampaka_2019, Villanueva-Domingo_2022}, by using one or the other simulation output. However, different galaxy simulations are, in the detail, realistic to different degrees and differently so in different regimes \citep[e.g.][]{2021MNRAS.501.4359Z}. It is therefore important to investigate how the various cosmological galaxy simulations, and hence the various sub-grid modeling choices included therein, can affect the inference of physical processes. Very few works have explored this issue until now, even though some momentum has been built recently with projects such as CAMELS \citep{CAMELS}, which has provided the community with a large dataset of galaxy simulations spanning a variety of standard cosmological parameters and of sub-grid choices around the fiducial ones advocated by the projects mentioned above.
Similarly, \cite{lovell} have coupled two different cosmological simulations, EAGLE and Illustris, with machine learning to infer the galaxy star formation histories from galaxy spectra. 

In this paper, we focus on the problem of deriving information about the merger histories of galaxies from observable properties, chiefly stellar observables. In fact, a key feature of the $\Lambda$CDM model is that galaxies evolve in a hierarchical fashion through successive mergers that consecutively lead to more massive structures \citep{evolution}. During this process, galaxies accrete mass from satellite galaxies through tidal stripping and merging: this contribution is often referred to as the ex-situ or accreted stellar mass of the galaxy \citep[e.g.][and references therein]{2016MNRAS.458.2371R}, to distinguish it from the stellar mass created through internal star formation (in-situ). Understanding the contribution of in-situ and ex-situ processes to the stellar mass assembly of a galaxy can shed light on its cosmic past, as the evolution of a galaxy is thought to be tightly coupled with its merging and assembly history, particularly in the early universe \citep{Springel_2005}. Whereas the past merger history of any galaxy cannot be observed, it is expected that it leaves 
significant traces on observable quantities such as stellar population properties and stellar kinematics.  

Observationally, there exists no direct mapping between observable galaxy properties and their stellar assembly history. Studies of the ex-situ stellar mass in galaxies are hence mostly based on semi-analytical models or cosmological simulations. For example, \cite{zoomin_1} utilized zoom-in simulations to examine the distribution of accreted mass and find that most massive galaxies are dominated by ex-situ stars. On the contrary, \cite{zoomin_2} reported lower accretion rates for galaxies in the low-mass regime using again zoom-in simulations. According to the Illustris and IllustrisTNG simulations, the ex-situ stellar mass fraction is highly correlated with stellar mass, in the sense that more massive galaxies have higher accreted mass fractions \citep{tng_exsitu1, 2018MNRAS.475..648P}. A similar relation was reported by \cite{eagle_exsitu}, who studied the ex-situ galaxy growth within the EAGLE cosmological simulation. 

Previous works have used ML to calibrate, i.e. quantitatively extract, the relation between observable properties and mergers by using the output of cosmological galaxy simulations. \cite{Bottrell_2019} utilized convolutional neural networks (CNNs) to classify mergers using different sets of simulated images to evaluate the importance of adding observational realism to the training data. \cite{Ferreira_2020} employed the IllustrisTNG simulations to train CNNs for the inference of major mergers up to $z\sim3$ and reported a good accuracy when applied on the CANDELS survey. \cite{Shi_2022} used a random forest approach to predict the ex-situ stellar mass fraction of galaxies from integrated halo and galaxy features from the TNG100 simulation, one of the IllustrisTNG flagship runs.

Here, we add to the ERGO-ML project (Extracting Reality from Galaxy Observables with ML), by building up and improving upon the work by \cite{Eisert_2022}. The latter showed with the IllustrisTNG simulations that it is possible to infer a number of summary statistics of the assembly history of galaxies from a small set of integral galaxy properties commonly available in large photometric surveys. In particular, conditional Invertible Neural Networks were used to predict the stellar ex-situ fraction, the average merger lookback time, the average merger stellar mass ratio, and the lookback time and stellar mass of the last major merger of simulated galaxies starting from their stellar mass, redshift, stellar half-light radius, stellar diskyness, integrated colour, average stellar metallicity, and average stellar age. However, both \cite{Eisert_2022} and all the works mentioned above used one only cosmological simulation model, for both training and testing.

In this work, we therefore proceed and quantify the impact of different simulations, i.e. different galaxy-formation models, when attempting to derive the merger histories of galaxies with ML. This is a fundamental step towards the application to observational data of ML models trained on simulations, a step that has not yet been tackled quantitatively. In particular, we use two state-of-the-art cosmological simulations, the aforementioned TNG100 and EAGLE, and we elect the galaxy-wide ex-situ stellar mass fraction as an indicative quantity of the past merger history of a galaxy. We aim at inferring it from two sets of observables: from a set of integral galaxy properties (similarly to what done by \citealt{Eisert_2022}) as well as from a set of two-dimensional maps representing the stellar content of galaxies. Although we do not include any forward modeling and observational realism in the construction of the spatially-resolved galaxy maps, these are simulation-based analogs to those that can be obtained with modern integral Field Unit surveys, such as MaNGA.
By producing simulated galaxy datasets created in the same manner from the two different simulations, we are able to cross-test across the two models and to investigate whether the trained neural networks have inferring power that surpasses the simulations' differences.  \looseness-2




The paper is organised as follows: in Section~\ref{sec:data} we briefly describe the simulations TNG100 and EAGLE and introduce the uniform dataset creation for both galaxy-property catalogs and 2D maps representations. The ML architecture details of the dense and the convolutional neural networks are supplied in Section~\ref{sec:method}. The initial performance of models is presented in Section~\ref{sec:models_eval}, where we quantify the accuracy of the models when evaluated on a not-seen-before test set as well as the domain drift observed when attempting to cross-test between simulations. In Section~\ref{sec:investigate_domain_drift}, we investigate the origins of this domain drift and propose a robust model across simulations in Section~\ref{sec:robust_model}. In Section~\ref{sec:discussion}, we further explore the underlying differences between the two cosmological simulations by attempting to understand what the models learn and we compare the most robust models across simulations with similar works. We conclude in Section~\ref{sec:conclusion}. \looseness-2

\section{Simulation data} \label{sec:data}

\subsection{Cosmological galaxy simulations}
We aim to create a robust neural network model able to infer the stellar ex-situ mass fraction of galaxies using only observable information. Since the merging history of a galaxy is not directly accessible from observational surveys, we turn to cosmological simulations where galactic evolution is tracked at distinct snapshots. We combine two state-of-the-art cosmological simulations, IllustrisTNG (TNG100) and EAGLE (Ref-L100N1504), described briefly as follows. 

\subsubsection{TNG100}
The IllustrisTNG Project\footnote{https://www.tng-project.org/} \citep{2018MNRAS.480.5113M, 2018MNRAS.475..676S, 2018MNRAS.475..648P, 2018MNRAS.475..624N, 2018MNRAS.477.1206N} is a series of magneto-hydrodynamic cosmological simulations that use the AREPO code \citep{springel_arepo}. Succeeding the original Illustris simulation, the TNG project was introduced with an updated galaxy-formation model \citep{2017MNRAS.465.3291W, 2018MNRAS.473.4077P} as well as refinements in physical processes. TNG consists of three primary runs with varying volumes and resolutions: TNG50, TNG100, and TNG300. In this work, we use the TNG100 simulation spanning a volume of $110^3 \rm{Mpc}^3$ with $1820^3$ dark matter particles and initial gas cells. The stellar mass resolution is  $\sim {10}^{6}\ \rm{M}_{\odot }$.

\subsubsection{EAGLE}
The EAGLE \citep{eagle, eagle2} is a suite of cosmological, hydrodynamical simulations developed by the Virgo Consortium\footnote{http://virgo.dur.ac.uk/}  designed to understand the formation and evolution of galaxies. The simulations were run using a
a modified version of the Gadget-3 Smoothed Particle
Hydrodynamics (SPH) code \citep{springel_gadget}. Here, we use the largest simulation volume, known as Ref-L100N1504 - hereafter referred to as EAGLE-L100 - with a volume of $100^3 \rm{Mpc}^3$. EAGLE-L100 uses $1504^3$ dark matter particles and an equal number of gas particles. The stellar mass resolution imposed is also $\sim {10}^{6}\ \rm{M}_{\odot }$, similar to that of TNG100.

\subsection{Measurement of the ex-situ fraction}
The first step towards creating the datasets from the two simulations is to identify the ground-truth we aim to infer, namely the stellar ex-situ mass fraction $\rm{f_{ex}}$ of a galaxy, in the same fashion for both simulations. For this calculation,  we need to measure how much stellar mass was created ex-situ for every galaxy by examining the origin of each stellar particle individually. 

We adopt the definition given in \cite{tng_exsitu1} for characterizing a stellar particle as either in-situ or ex-situ. According to this work, in-situ particles are considered to be all particles that were formed in the main progenitor branch of the galaxy they currently reside in. Particles forming outside the main progenitor branch are considered ex-situ stellar particles.  The total ex-situ mass for each galaxy is calculated by aggregating the mass of stellar particles formed ex-situ. 

To ensure that the ex-situ stellar mass fractions are homogeneously calculated among TNG100 and EAGLE-L100, we decide to adopt a single identification scheme of structures (i.e. halo and subhalo finders) for both simulations. This way, we can remove any systematic variations caused by diverging definitions of the merging trees and we can entirely focus on capturing the effect of the different underlying sub-grid recipes on the derivation of merging histories. To that end, we decide to use a version of the EAGLE-L100 that has been analysed exactly as TNG100 \citep[details can be found in the data release paper][]{2019ComAC...6....2N} and has already been used in other works for isolating the impact of sub-grid variations on various physical processes \citep[e.g.][]{Ayromlou_2022}.

Following the TNG100 scheme, the merger trees are constructed with the SUBLINK algorithm \citep{2015MNRAS.449...49R}. A unique descendant is assigned to each subhalo on every snapshot by taking into account the number of shared particles between galaxies to identify the one more closely related between snapshots. Then, the progenitor tree with the most massive history of each subhalo is defined as the main progenitor branch. It is worth noting that the merger trees are constructed by taking into account only baryonic particles, in particular stellar particles and star-forming gas cells of subhaloes. By traversing the merging tree of a subhalo, one can resolve the origin of the stellar particles currently residing in this galaxy. \looseness-2

Luckily, for TNG100 there already exists a catalog for every snapshot that classifies each particle either as ex-situ or in-situ \citep{2015MNRAS.449...49R, tng_exsitu1, 2017MNRAS.467.3083R}. In the same work, the total in-situ and ex-situ stellar mass is also calculated in an aggregated catalog. As a result, measuring the ex-situ mass fraction for our TNG100 dataset is straightforward. Similar catalogs are also available for the custom analysis of the EAGLE-L100 run, that have been calculated with the same procedure. Hence, we are able to create our datasets of ex-situ stellar mass fraction in an identical fashion for both simulations. \looseness-2

\subsection{Galaxy sample selection}

In order to create datasets that are directly comparable between TNG100 and EAGLE-L100, we use the same specifications for both simulations. The sample of galaxies that is selected is of stellar mass $> {10}^{10}\ {M}_{\odot }$. This selection emerges from the fact that fewer particles exist on the simulations for lower mass galaxies, thus rendering the creation of two-dimensional maps rather difficult and uninformative.\looseness-2

Originally we consider only subhalos at redshift $z = 0$, but we find that both galaxy samples are highly imbalanced, as the low-end of ex-situ mass fraction galaxies prevails and there exist very few galaxies with a high ex-situ mass fraction $(> 0.7)$. This can have a major effect on the accuracy of the neural network predictions. To address this issue, we include more snapshots in our dataset. More specifically, we update our EAGLE-L100 dataset with subhalos with an ex-situ mass fraction $f_{ex} > 0.2$ from redshifts $z = 0.1$ and $z = 0.2$. Accordingly, we update our TNG100 dataset with subhalos with an ex-situ mass fraction $f_{ex} > 0.2$ from redshift $z = 0.1$.

\begin{figure}
    \centering
    \subfloat[Original sample  used for the integrated properties approach]{\includegraphics[width=0.5\linewidth]{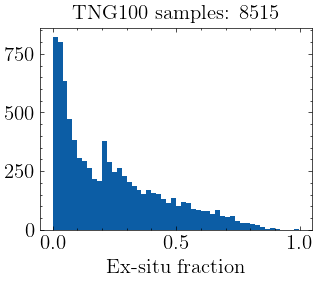}\includegraphics[width=0.5\linewidth]{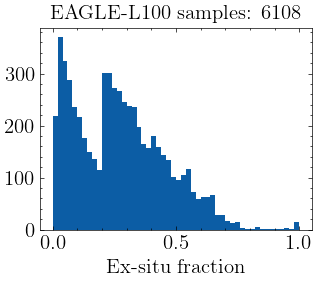}}
    \\
    \subfloat[Balanced sample used for the 2D spatial maps approach]{\includegraphics[width=0.5\linewidth]{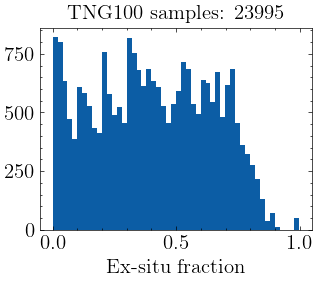}\includegraphics[width=0.5\linewidth]{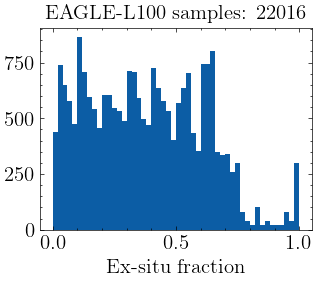}}
    \caption{The distribution of the ex-situ stellar mass fraction values for the selected samples from the TNG100 and EAGLE-L100 simulations. We select all galaxies with stellar mass $> {10}^{10}\ {M}_{\odot }$ from $z = 0$ (TNG100 and EAGLE-L100) and galaxies of the same stellar mass range with an ex-situ stellar mass fraction $f_{ex} > 0.2 $ from $z = 0.1$ (TNG100 and EAGLE-L100) and $z = 0.2$ (EAGLE-L100), in an initial attempt to balance our dataset samples. \textbf{Top:} Histogram of ex-situ stellar mass fractions of the original TNG100 (on the left) and EAGLE-L100 galaxy samples (on the right). This sample distribution is used for the integrated values inference approach. \textbf{Bottom:} Histogram of ex-situ mass fractions of the two simulation samples after balancing by using multiple projections per galaxy. This sample is used for the 2D spatially-resolved maps inference approach.}
    \label{fig:balancing_ds}
\end{figure}

The final sample from TNG100 consists of 8515 galaxies. Respectively, our final sample from EAGLE consists of 6108 galaxies. From these samples, we will create two different input datasets to describe the observable realisations of a galaxy, the former containing only integrated values of observable properties and the latter their respective 2D spatially-resolved projection maps. A construction advantage of the 2D maps approach is that we can further balance our datasets as well as increase our training sample size by utilizing more than one projection per galaxy, as it will be further discussed in \ref{2d_maps}. The distribution of the ex-situ stellar mass fractions of the original sample for TNG100 and EAGLE-L100 can be found in Figure \ref{fig:balancing_ds} (a). The balanced samples utilized for the 2D spatial maps approach are respectively illustrated in Figure \ref{fig:balancing_ds} (b). We proceed by describing the procedures followed to create the two datasets. 

%

\subsection{Observable integrated properties}

Our initial approach is to investigate whether we can infer the ex-situ stellar mass fraction of galaxies from integrated values of some of their fundamental properties, that can be also be obtained from observations. Additionally, we wish to understand if such properties are concise reflectors of the merging history across simulations, independent of the different sub-grid recipes. To that end, we choose the following properties that will be used as inputs:

\begin{itemize}
    \item Stellar Mass
    \item Mass-weighted stellar metallicity
    \item Mass-weighted stellar age
    \item Half-mass stellar radius
    \item The total spin of the galaxy
\end{itemize}

We note here that, while the total spin of the galaxy is not a directly observable property, it can be indirectly inferred from observations. A clear correlation with the observable spin of the stellar component has been shown, especially for late type galaxies \citep{2022MNRAS.512.5978R}. We therefore include it as one of the integrated properties considered as inputs.

In Table \ref{tab:integral_inputs_table}, a brief description of the integrated properties from both simulations is shown as well as which fields were used to extract them. We proceed by comparing the distributions of the integrated properties between the two simulations to check how well the two samples are matched and if they exhibit a relation to the value we wish to infer, namely the ex-situ mass fraction. This comparison can additionally provide an initial insight on how independent the ex-situ stellar mass fraction relations are from the different sub-grid recipes between simulations.

\begin{figure*}
    \centering
    \includegraphics[width=\textwidth]{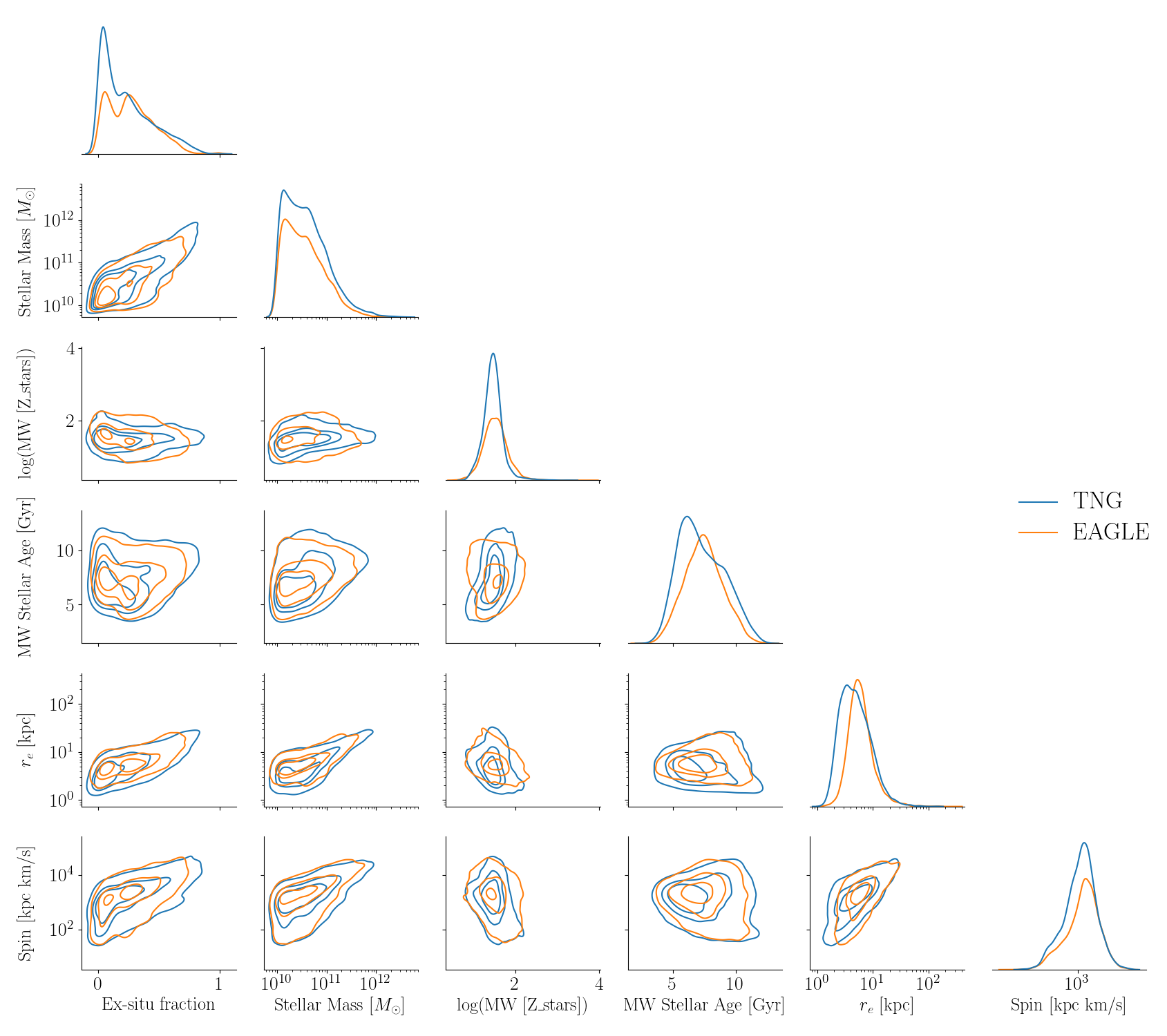}

    \caption{Comparison of relations between ex-situ stellar mass fraction versus the stellar mass, mass-weighted metallicity (log), mass-weighted age, half-mass stellar radius and spin of galaxies from the TNG100 (blue contours) and EAGLE-L100 sample (orange contours). The diagonal displays the distribution of the properties on focus for both simulations. The entire dataset samples of galaxies from both simulations are utilized for this plot (${M}_{*} > {10}^{10}\ {M}_{\odot }$ from redshifts $z = 0, 0.1, 0.2$). We find a good match between the two simulation samples, which will probably allow our trained models to infer the ex-situ stellar mass fraction from the integrated properties across simulations.}
    \label{fig:properties_comp_corner_plot}
\end{figure*}

In Figure \ref{fig:properties_comp_corner_plot}, we show how the ex-situ mass fraction relates with the stellar mass, the metallicity, age, stellar half-mass radius ($r_{e}$) and spin of the galaxies in the TNG (blue contour) and EAGLE (orange contour) datasets. Furthermore, the histogram of each property is shown in the diagonal for both simulations. We find that the two simulations seem well-matched in all properties. Only slight differences exist; for instance the TNG sample contains relatively younger and more compact galaxies.

As far as the relations with the ex-situ stellar mass fraction are concerned, we note the same trends appear in both simulations. The ex-situ mass fraction is highly correlated with the stellar mass, in the sense that more massive galaxies have higher ex-situ fractions in both simulations. The same correlations are present for the galactic size and spin. We anticipate that this agreement will allow the neural network models to infer across simulations when we use as input the integrated properties values.

\subsection{Observable 2D spatially-resolved maps} \label{2d_maps}

The second approach we propose for inferring the ex-situ stellar mass fraction of galaxies is through 2D projections of their observable properties. To differentiate further from the previously described integrated inputs, we decide to remove correlations that exist between the inputs and the ex-situ fraction and use only gradients present in the spatial maps, which are likely more linked to accretion histories and possibly less biased towards a specific subgrid recipe, as it will be discussed in section \ref{remove_dep}.

\subsubsection{Spatial maps creation}

\begin{figure*}
    \centering
    \subfloat[]{\includegraphics[width=\textwidth]{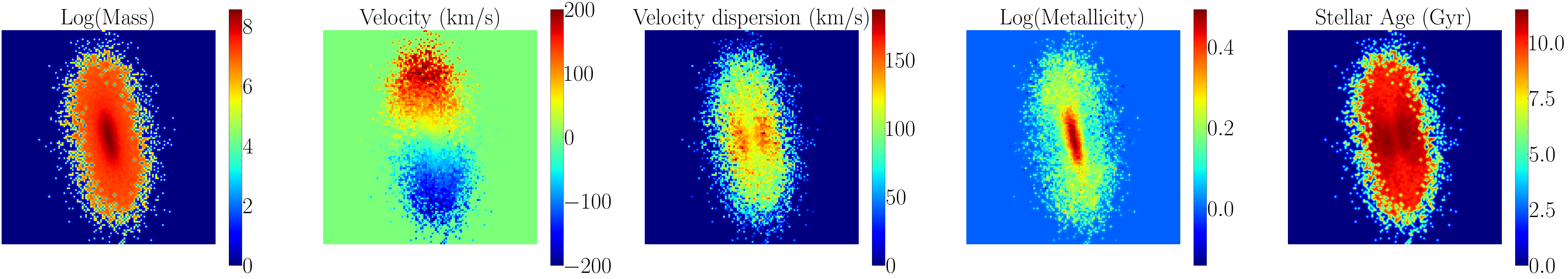}}\\
    \subfloat[]{\includegraphics[width=\textwidth]{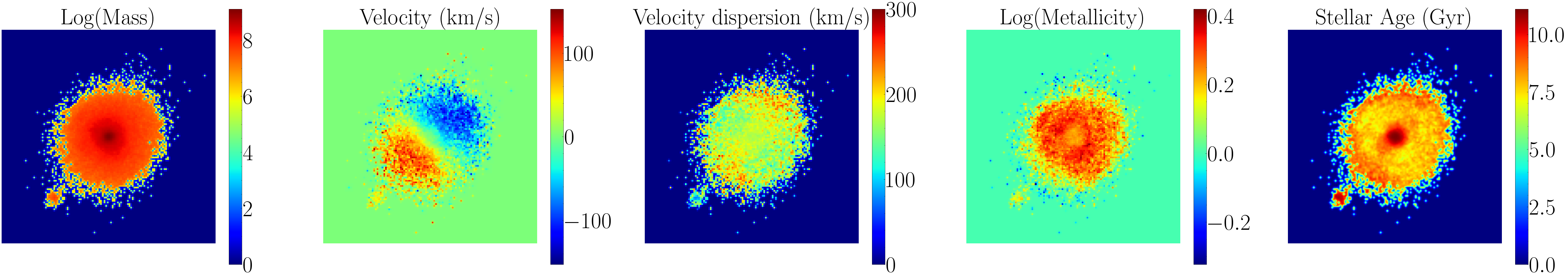}}
    
    \caption{The two-dimensional spatially-resolved maps from two sample galaxies from the TNG100 and EAGLE-L100 simulations, which will be used as inputs for inferring the ex-situ stellar mass fraction through neural networks. \textbf{Top:} Two-dimensional maps of the spatial distribution of stellar mass, velocity, velocity dispersion, metallicity (log) and age for a sample galaxy from the TNG100 simulation with ex-situ stellar mass fraction $f_{ex} = 0.28$.  \textbf{Bottom:} Two-dimensional spatial maps for a sample galaxy from the EAGLE-L100 simulation with ex-situ stellar mass fraction $f_{ex} = 0.53$.}
    \label{fig:2d_maps}
\end{figure*}

In this section, we proceed by describing how the two-dimensional observable spatial maps are created. We choose to extract noiseless maps from the simulations, without adding extra realism, so that we can evaluate the predictive power in idealized conditions.
For each galaxy in our original sample, we produce the two-dimensional spatially-resolved maps by projecting the stellar particles along a given line-of-sight. The aperture is selected as four times its stellar half-mass radius ($r_{e}$) across the image and the resulting spatial maps have a fixed size of 128 x 128 pixels. This implies that the pixel scale is not fixed but instead dependent on the $r_{e}$ of each galaxy. \looseness-2

We select five different kinds of observable spatial maps to predict the ex-situ stellar mass fraction. The first one is the stellar mass density map, as it is known that the stellar assembly history of a galaxy is closely related to its the stellar mass.

Additionally, we wish to study whether adding the kinematic information of the stellar populations can enhance the prediction of the ex-situ stellar mass fraction, as kinematics are known to be directly affected by mergers \citep{emsellem_2007, krajnovi_2011, grad_2017}. To this end, we include maps of the spatial distribution of velocity and velocity dispersion to our observable set.  \looseness-2

Finally, we speculate that stellar population properties of a galaxy might prove a significant reflector of its merging history \citep{navarro, rossi, age_gradients}. Therefore we add 2D spatial maps of mean stellar metallicity and mean stellar age to the observable dataset. A brief description on which fields are utilized from both simulations as well as the binning operation can be found in Table \ref{tab:2d_maps_table} in the Appendix.

We create the 2D observable spatial maps by projecting all stellar particles of the galaxy along a given line-of-sight. In order to create multiple realisations per galaxy that can serve as independent observations, we place cameras at distinct positions to observe the galaxy along a unique line-of-sight. To ensure that each camera is able to capture different morphological features of the galaxy, we place a regular polyhedron at the gravitational potential minimum of the subhalo and use its vertices as the distinct locations of the cameras. A similar approach has been used in \cite{Bottrell_2017, Bottrell_2017_ii, Bottrell_2019}. We do not include observational effects since we want to explore the effect of model mispecification.

Instead of creating a fixed number of projections per galaxy, we decide to vary that number by taking into account their stellar ex-situ mass fraction. This way, we can enable the balancing of our dataset, as we can dictate the number of realisations produced per galaxy by selecting a different polyhedron based on the number of vertices and by extension the possible line-of-sights it offers. More specifically, we produce a higher number of projections for galaxies with a high ex-situ fraction to account for their low number densities and a lower number of projections for galaxies with a low ex-situ fraction. Each line-of-sight is then considered as an independent observation. For our setup, we utilize the following regular polyhedra:

\begin{itemize}
    \item tetrahedron (4 projections)
    \item octahedron (6 projections)
    \item cube (8 projections)
    \item icosahedron (12 projections)
    \item dodecahedron (20 projections)
\end{itemize}

In an attempt to denoise the observable spatially-resolved maps, we choose to ignore bins with a low number of stellar particles. The threshold on the number of stars is selected individually for each galaxy projection. More specifically, we filter out the 75 per cent of all pixels and keep only the 25 per cent with the highest amount of stellar particles, by taking into account also the pixels belonging to the background.

A set of the observable spatial maps of a sample galaxy from TNG100 in a random 2D projection can be found in Figure \ref{fig:2d_maps} (a). Correspondingly, the spatial maps of a EAGLE-L100 galaxy are shown in Figure \ref{fig:2d_maps} (b). \looseness-2

The final TNG100 and EAGLE-L100 datasets consists of 23995 samples and 22016 samples respectively. In Figure \ref{fig:balancing_ds} (b), the distribution of stellar ex-situ mass fractions for both datasets is illustrated. It is apparent that the datasets, while being more or less balanced for values $< 0.8$, still suffer from very few samples at high ex-situ fractions. However, they still hold a balancing advantage over the respective original sample used on the integrated inputs approach (Figure \ref{fig:balancing_ds} (a)). \looseness-2

\subsubsection{Removing mass and size dependency} \label{remove_dep}

We aim to develop two approaches for inferring the ex-situ stellar mass fraction of galaxies that take advantage of different features present in the inputs. We have already demonstrated that the ex-situ stellar mass fraction of a galaxy is correlated with its stellar mass. More specifically, more massive galaxies tend to have a higher ex-situ stellar mass fraction in general and vice versa. Additionally, as shown in Figure \ref{fig:properties_comp_corner_plot} the ex-situ stellar mass fraction also depends on the size of the galaxy and there exist further trends.

Since this information will most likely be used from the neural networks trained on the integrated properties and the dependence might differ between simulations, we attempt to remove the size and mass correlations from our 2D datasets. This is based on the idea that gradients represent a less sub-grid dependent record of the past merging history of galaxies.

To that end, we take two decisions. First, we scale all galaxies to a fixed aperture that is equal to four times the half stellar mass radius of each galaxy. This way the size dependency is removed, as all galaxies are viewed from at the same scale and the pixel scale is dependent on each galaxy. This decision is enforced on the creation of all the observable spatially-resolved maps from the simulation particles.

\begin{figure*}
    \centering
    \subfloat{\includegraphics[width=\linewidth]{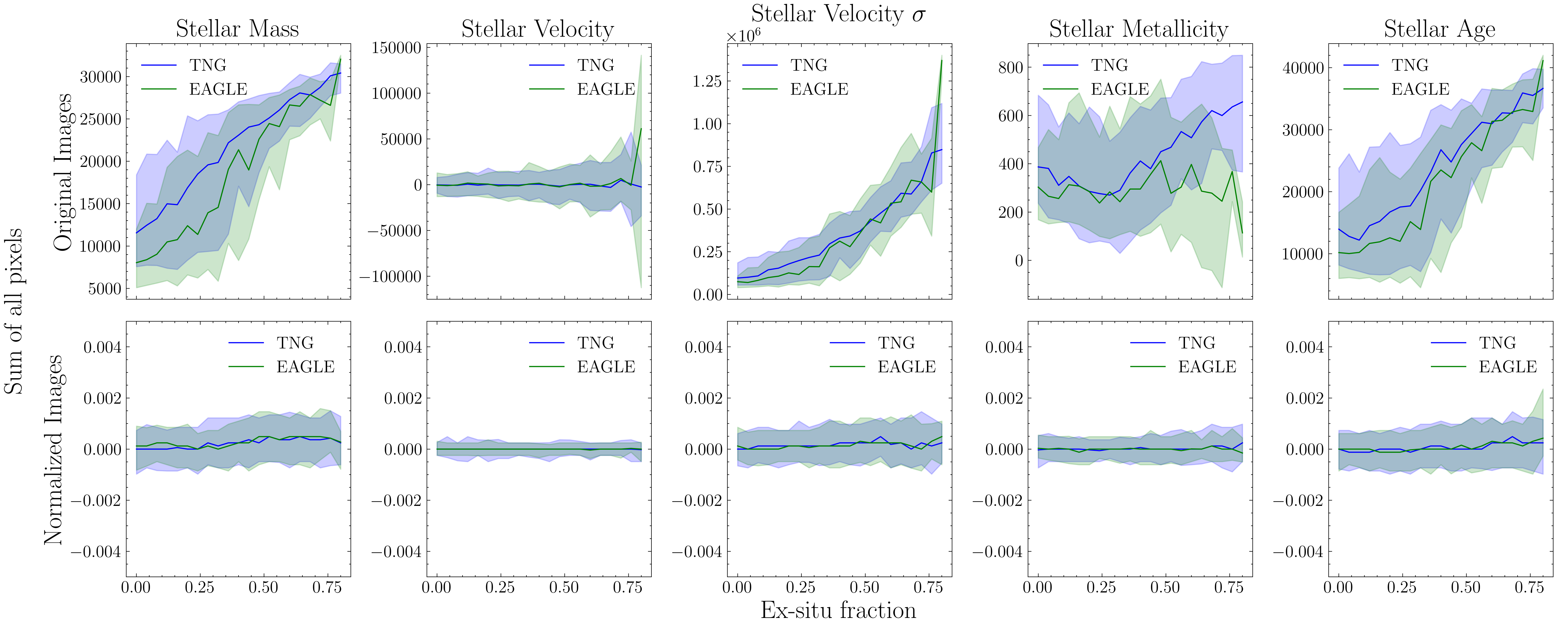}}
    \caption{Trends existing in the 2D spatially-resolved observable maps of the TNG100 and EAGLE-L100 simulations with the ex-situ stellar mass fraction. We plot the sum of all pixels in a respective 2D map versus the ex-situ stellar mass fraction for all galaxies in our datasets. The solid lines represent the median value and the shaded regions contain 68 per cent of the data. \textbf{Top:} The trends between the sum of all pixels in a 2D spatial map and the ex-situ stellar mass fraction are shown for the original unprocessed 2D spatial maps. \textbf{Bottom:} The same relations are illustrated but on the normalized images of the spatial distributions, after the standardization per map has been applied. While the original 2D spatial maps display several strong correlations with the ex-situ fraction, we remove all these trends using the selected normalization scheme and thus focus only on the gradients present on the spatial maps.}
    \label{fig:trends}
\end{figure*}

Additionally, we remove the mass dependency and all other trends by standardizing every observable map individually. In Figure \ref{fig:trends}, the relation between the sum of all pixel values of each respective observable spatial map in relation to the ex-situ mass fraction of the galaxy is illustrated. It is apparent that there exists a strong correlation between this aggregated value and the ground-truth we wish to infer on the mass, velocity dispersion and age observable spatial maps. This is anticipated, since more massive galaxies have a higher velocity dispersion and are generally consisting of older more massive stars. 
A weak relation also exists between the metallicity and the stellar ex-situ fraction.  \looseness-2

By standardizing each observable map individually, we remove this correlation as it is illustrated on the bottom row of Figure \ref{fig:trends}. Since every map is standardized so that it includes values with a mean of 0 and a standard deviation of 1, all mean values are close to zero. This way, the model is now forced to extract features related to the shapes present in the observable spatial maps and not their absolute values.

\section{Methods}  \label{sec:method}

\subsection{Model architecture}

We wish to infer the ex-situ stellar mass fraction of galaxies from two different sets of inputs, namely summary statistics of their observable properties and 2D spatially-resolved maps of observable quantities. Since our dataset derives from simulations where the ground truth is available, a supervised machine learning approach is considered suitable for this task. Additionally, in order to be able to measure the confidence of the model, we choose a probabilistic regression architecture that predicts a posterior probability distribution for each input.

In particular, we use Mixture density networks (MDNs) \citep{Bishop1994MixtureDN}, which are a specific family of networks that capture the aleatoric uncertainty of the model by combining a conventional neural network with a probability mixture model. The output of a MDN is a probability distribution for each input that can be a mixture of distributions. After training, one can sample from the posterior produced when applying the model on the unseen test data to calculate the uncertainty of the predictions.

Since we have two different kinds of inputs, namely a set of summary values and 2D images, we need two separate neural network architectures to handle them. For the integrated inputs, we choose a simple multilayer perceptron network (MLP). For the 2D images, we select a convolutional neural network (CNN). 

Both model architectures are built with a MDN configuration, meaning that they receive the observables as inputs and after training, they estimate a posterior distribution that we assume to be a 1-component Normal. This is a strong assumption, but we find that adding more than one Gaussian distributions did not have a changing effect on the results. By sampling multiple times from this distribution, we can infer a mean value between one and zero corresponding to the galaxy's predicted ex-situ mass fraction. Additionally, the standard deviation of the posterior is the measure of uncertainty of the model on the particular prediction. 

The specific details of the architectures of the probabilistic MLP and the CNN can be found in the Appendix in Table \ref{tab:neural_network_dense} and Table \ref{tab:neural_network_cnn} respectively. As illustrated there, the two networks share a similar configuration, being different in the inputs they receive and the dense versus the convolutional layers. While the MPL has three simple dense blocks, the CNN has three convolutional blocks that each consists of four layers: convolution, max pooling, batch normalization and dropout. The convolution layers have 32, 32 and 64 filters respectively and a rectified linear unit (ReLU) activation \citep{relu} is utilized in all of them. Max pooling layers \citep{maxpooling} are placed after the convolution layers to enhance the model's invariance to small translations in the internal representation of the input. To optimize training, we add batch normalization layers. Finally, we use dropout layers as a measure to reduce overfitting. \looseness-2


We use the negative log-likelihood as the loss function of both model architectures:
\begin{equation}
    \centering
    \mathcal{L} = - \frac{1}{N} \sum \log{p(y|x)}
\end{equation}

Minimization of the loss function is done using the Adam optimizer \citep{adam_opt} with a learning rate of $1e-3$. The learning rate defines how much the weights are updated when performing the gradient descent. We train the MLP and the CNN for 500 epochs on their respective training sets, while additionally monitoring the loss on the validation set in order to avoid overfitting.

All model architectures are implemented in Keras \citep{chollet2015keras} with TensorFlow backend \citep{tensorflow2015-whitepaper} and using TensorFlow-Probability. To optimize the execution time, we train our models on a NVIDIA Tesla P100 GPU.


\subsection{Preprocessing}

Prior to training, we apply a number of preprocessing steps to the datasets to ensure that training is optimized.\looseness-2

\subsubsection{2D spatial maps refinement}

We initially wish to refine the 2D spatially-resolved observable maps since they originate from raw particle data from the simulations. We choose to apply the logarithmic function to the stellar mass and metallicity maps to better capture the structures present in the spatial maps. Additionally, we clip values above a specific threshold that is typical for this kind of observable maps. This is a precaution step to remove unreasonable values from simulations. More specifically, the spatial maps of velocity are clipped to values between $(-250, 250)$ $\rm{km~s}^{-1}$, the velocity dispersion spatial maps are forced to an upper limit of $350$ $\rm{km~s}^{-1}$ and the metallicity spatial maps are clipped between $(-1, 1)$ dex.

Finally, in an attempt to smooth the spatially-resolved maps, we apply a Gaussian filter to each map individually using a Gaussian kernel with standard deviation $\sigma = 0.6$ pixels. This simulates the effect of seeing in real observations. \looseness-2

\subsubsection{Feature scaling}

Feature scaling is a standard preprocessing step in most machine learning applications used to normalize the range of the data features. In our case in particular, it is a required procedure to ensure that all 2D spatial maps are treated uniformly by the neural network model irrespective of the data range they cover. In the integrated values approach, a standard scaler is utilized for each property to guarantee that the network places equivalent importance on all inputs.

\subsubsection{Sample splitting}

We randomly split our two datasets into a training (75 per cent), a validation (10 per cent) and a test subset (15 per cent). The training set is used to fit the weights of the model whereas the validation set is utilized during training to evaluate the performance of the model and prevent over-fitting. We measure the generalization performance of the trained models by applying them on a test set that consists of galaxies the network has not seen before.

In the 2D spatially-resolved maps approach, extra caution is taken across splitting to ensure that all projections of one galaxy are gathered in the same dataset type, so that to avoid leaks of information when testing the prediction results of our model. However, it should be noted that the same care is not applied on descendants and progenitors of a galaxy in different snapshots. We treat galaxies found in the same main branch as unique observations varying in morphological features. We consider that this is a safe choice that does not affect the evaluation
as the snapshots included are separated by hundreds of Myrs.

\subsubsection{Data augmentation}

 We augment our 2D data by enforcing a random flip down/up and left/right on each training step. The goal is to increase the variability of our dataset by creating different realisations of galaxies while maintaining their semantic information.

\section{Models evaluation} \label{sec:models_eval}

In this section, we collect the results of the initial evaluation of the models in terms of their accuracy in inferring the ex-situ stellar fraction of galaxies from observables across simulations. We train a variety of MLP and CNN models for the integrated input values and the 2D spatial maps approaches respectively and evaluate them on an unseen before test set. We choose to keep the same test sets for both simulations for all the different runs so that the results we acquire are directly comparable.

We investigate two different training setups:
\begin{itemize}
    \item Training on a set of galaxies from TNG100
    \item Training on a set of galaxies from EAGLE-L100
\end{itemize}

For each setup and each input approach, five different models are trained with identical architectures, different only on their weight initialization. We test each model on the both the TNG100 and the EAGLE-L100 test sets.

We acquire the predictions of each training setup on the test set by ensembling the five trained model results.  More specifically, since each model predicts a distribution when evaluated on a specific galaxy, we sample from all five inferred distributions multiple times for every galaxy. We then ensemble the predicted posterior from all samples. The ex-situ stellar mass fraction is predicted as the Maximum A-Posteriori value (MAP) of the ensemble distribution, while the produced standard deviation corresponds to the confidence of the models on each prediction. In section \ref{uncertainty} we investigate whether the standard deviation can serve as a meaningful measure for capturing the uncertainty of the model.

\subsection{Training and inference on the same simulation} \label{ground_truth}

The main purpose of building a neural network is to predict the ex-situ stellar mass fraction as close to the true value as possible. To this end, the primary evaluation metric of the performance of the models is their prediction accuracy.

As previously discussed, we employ two different input schemes to infer the ex-situ stellar mass fraction, namely a set of integrated galaxy properties and 2D map projections.
We wish to quantify how well each approach performs on the two simulation test sets as well as if a domain drift is observed between the TNG100 and EAGLE-L100 simulations. To that end, we initially attempt to predict the ex-situ stellar fraction by training on the same simulation that we wish to test upon, which is bound to produce relatively good results. 

\begin{figure}
    \centering

    \large\text{Evaluation on the TNG100 test set}\par\smallskip

        \subfloat[Predict with catalog values]{\includegraphics[width=0.5\linewidth]{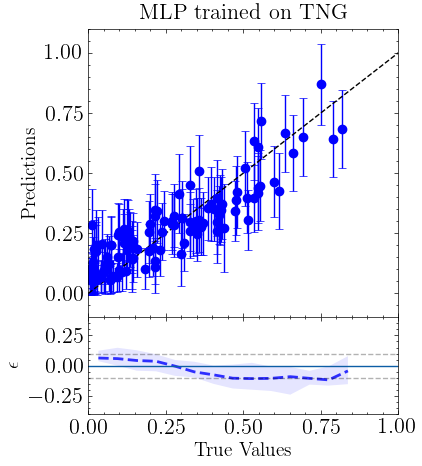}}
        \subfloat[Predict with 2D spatial maps]{\includegraphics[width=0.5\linewidth]{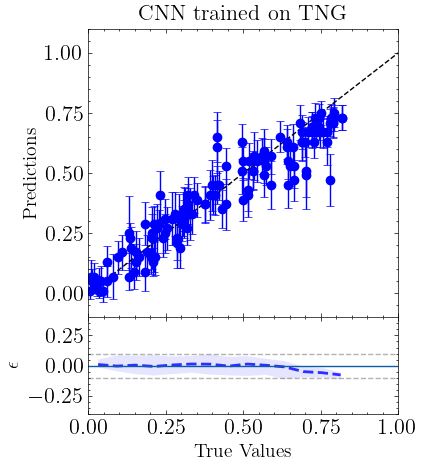}
        }
        
    \vspace{0.5cm}
    
    \text{Evaluation on the EAGLE-L100 test set}\par\smallskip

        \subfloat[Predict with catalog values]{\includegraphics[width=0.5\linewidth]{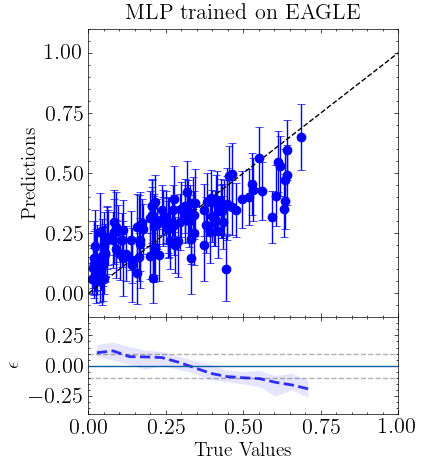}}
        \subfloat[Predict with 2D spatial maps]{\includegraphics[width=0.5\linewidth]{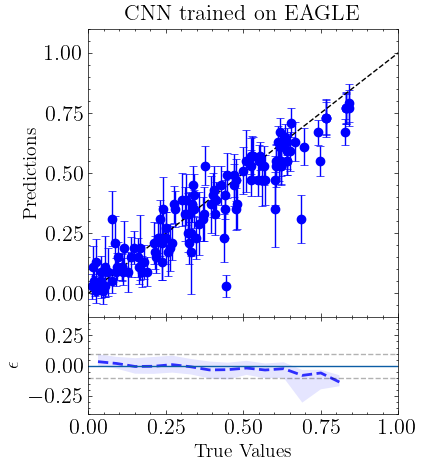}}

    \caption{Evaluation of models trained on a fixed simulation scenario with the integrated inputs (a, c) and the 2D spatial maps (b, d) for the TNG100 (upper row) and EAGLE-L100 (bottom row) test sets. In each panel, the ensemble predictions of the models versus the ground truth of the ex-situ stellar mass fraction is illustrated for 128 galaxies randomly selected from the two test sets. The error bars correspond to the standard deviation of the ensemble posterior that is predicted for each galaxy. The black dashed line on the prediction panels marks the 1:1 relation and serves as a guide to the eye, as the prediction points should be gathered as close to that line as possible. On the bottom error panels, the blue dashed line indicates the median of the error in the prediction, i.e. the difference between the model's estimate and the ground truth. The shaded region contains the 68 per cent of the data.}
    \label{fig:tng_eagle_test_set_results}
\end{figure}

In Figure \ref{fig:tng_eagle_test_set_results}, we collect the results of the trained MLPs and CNNs on the TNG100 test set (upper row) and the EAGLE-L100 test set (lower row). For each test set, we show both the catalog predictions (a, c) and the 2D map predictions (b, d) on the respective test set. 

We initially evaluate the performance of the models on the TNG100 test set (upper row of Figure \ref{fig:tng_eagle_test_set_results}). The MLP models trained on the integrated input values demonstrate a relatively good predictive power when tested on the test set originating from the same simulation (a). However, this inferring power deteriorates on higher ex-situ fractions, as bias goes up to 0.1 for all median and high ex-situ fractions. On the contrary, CNN models trained on the 2D spatial input maps (b) show a very high predictive power that remains strong on all ex-situ regimes. The error between the predictions and the ground-truth is always well-bounded between (-0.1, 0.1) with a very low mean bias (< 0.02) and the scatter is much lower compared to the one produced by the respective integrated prediction models.

We proceed by evaluating the inference results on the EAGLE-L100 test set in the same fashion, as illustrated on the bottom row in Figure \ref{fig:tng_eagle_test_set_results}. We now find a much larger scatter and a higher bias even on the single simulation scenario from MLP models (c). Particularly, on higher ex-situ stellar mass fractions the prediction errors reach absolute values that go up to 0.25. Contrarily, the CNN models (d) demonstrate a superiority on accuracy, that is well propagated on all ex-situ regimes. \looseness-2

Overall, we find that the prediction through the integrated values in a single simulation scenario is introducing a higher bias especially on the EAGLE-L100 simulation and particularly on higher ex-situ stellar fractions. This can be attributed to the significant imbalance of the original datasets. This imbalance can be alleviated in a 2D map input approach, even if the same galaxy sample is used, since we can construct multiple projections per galaxy, thus additionally increasing our dataset size. We find a significantly lower bias as well as a lower scatter in the predictions from the trained CNNs on the 2D spatial maps in a single simulation setup. \textit{We underline here that this results suggests that the information present in the shapes/gradients of observable spatial maps is sufficient for inferring the unobservable ex-situ stellar mass fraction.}

\subsection{Training and inference on different simulations}\label{ground_truth_cross}

\begin{figure}
    \centering

    \large\text{Cross-evaluation on the TNG100 test set}\par\smallskip

        \subfloat[Predict with catalog values]{                \includegraphics[width=0.5\linewidth]{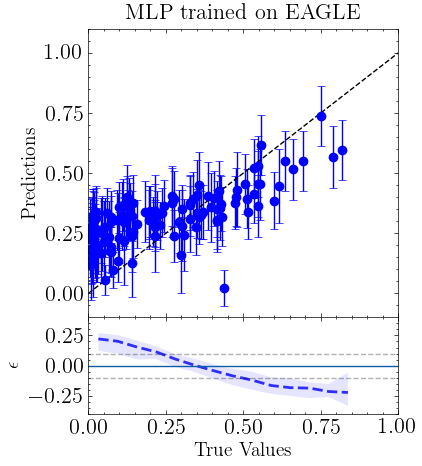}}
        \subfloat[Predict with 2D spatial maps]{
        \includegraphics[width=0.5\linewidth]{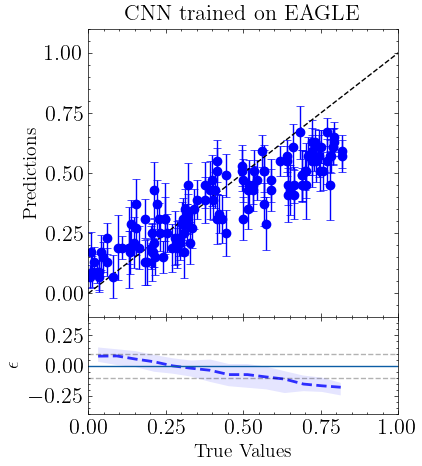}}
        
    \vspace{0.5cm}
    
    \text{Cross-evaluation on the EAGLE-L100 test set}\par\smallskip

        \subfloat[Predict with catalog values]{\includegraphics[width=0.5\linewidth]{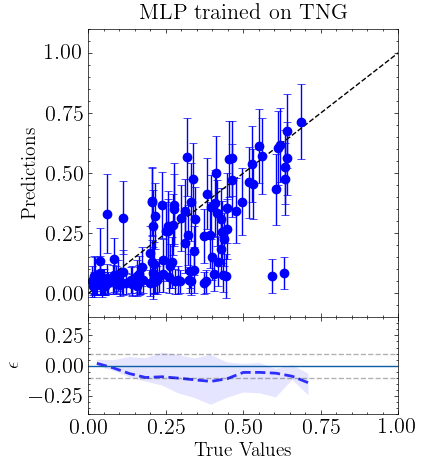}}
        \subfloat[Predict with 2D spatial maps]{\includegraphics[width=0.5\linewidth]{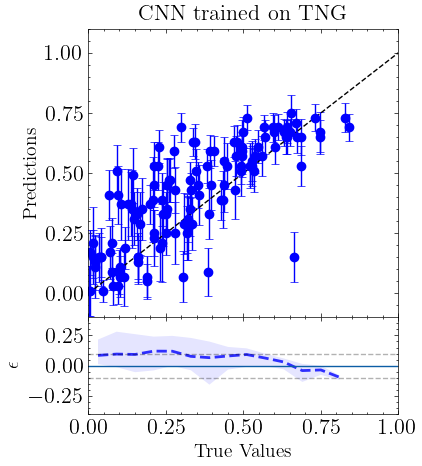}}

    \caption{Cross-evaluation of models trained on a different simulation than the test set with the integrated inputs (a, c) and the 2D spatial maps (b, d) for the TNG100 (upper row) and EAGLE-L100 (bottom row) test sets. Similar to Figure \ref{fig:tng_eagle_test_set_results}, we illustrate the predictions versus the ground-truth along with the error bars, as they were predicted from models trained across simulations. We find a general deterioration in the predictions compared to the fixed simulation scenario.}
    \label{fig:tng_eagle_test_set_results_cross}
\end{figure}

We proceed by cross-testing the trained neural networks to evaluate if our models have a predictive power that surpasses the sub-grid physics variations across simulations. More specifically, we evaluate the performance of models that have been trained on EAGLE-L100 on the TNG100 test set and vice versa. This is a trickier procedure, since the models might not generalize well across simulations. \looseness-2

In Figure \ref{fig:tng_eagle_test_set_results_cross}, we collect the obtained results from the cross-tests. Initially, we evaluate the predictive power of models trained on EAGLE-L100 on the TNG100 test set. The MLP models trained on EAGLE-L100 (a) demonstrate a declining prediction accuracy, particularly on higher and lower ex-situ fractions (bias goes up to 0.25). The corresponding cross-test from the CNNs trained the 2D spatially-resolved maps (b) reveals a lower prediction bias than the catalog approach. However, there still exists a significantly larger scatter in comparison to the fixed simulation scenario and an apparent under-prediction in the higher ex-situ regimes. This is partly anticipated since the two simulations vary substantially in their cosmological recipes and the training might not be able to surpass these differences. Nonetheless, training a model on EAGLE-L100 and testing on TNG100 still has a relatively strong prediction power on the 2D spatial maps approach. This indicates that the training sample originating from EAGLE-L100 is in its majority well represented in the TNG100 dataset. \looseness-2

On the other hand, the prediction power of both the MLP and the CNN models rapidly decreases when the EAGLE-L100 test set is evaluated on models cross-trained on TNG100. A very high bias (> 0.25) and scatter is introduced from the MLP models (c). Using the 2D spatially-resolved input maps (d), the predicted ex-situ fraction of galaxies residing at the lower end is largely overestimated, with a bias up to 0.1 for all galaxies with an ex-situ fraction $<0.6$. This suggests that the models trained on TNG100 do not generalize well on EAGLE-L100. A possible cause of that would be that there exist stellar populations on TNG100 that are not present on EAGLE-L100. As a result, the CNN learns more complex features from the observable spatial maps that fail to infer on EAGLE-L100, as these features might not be present. We investigate this in more detail in section \ref{channels}. \looseness-2

In summary, upon attempting to cross-test our models across simulations we find that a higher bias in the predictions and a larger scatter is introduced in both input approaches. This domain bias might be propagated in a possible application to actual observations. Attempting to explain this domain drift and investigate how to overcome it, we focus our work henceforward on the 2D spatial maps input approach. There, the predictive power in a single simulation scenario is much higher and the domain drift is still evident across simulations. \looseness-2

\section{Investigating the domain drift} \label{sec:investigate_domain_drift}

\subsection{Importance of Channels in 2D maps} \label{channels}

We wish to better understand which spatilly-resolved maps hold more information regarding the inference of the ex-situ stellar mass fraction. It is possible that the gap between the effect of including a certain input map versus another is so significant, that we could omit the latter altogether. Also, we aim to test whether including only a subset of the spatial input maps can still lead to an efficient inference.

More importantly, we are keen to explore whether the inclusion of a certain type of maps in the input of the CNN impedes the inference across simulations. It was noted on section  
\ref{ground_truth_cross}, that while all CNNs are very efficient in predicting the ex-situ stellar mass fraction when tested on a set originating from the same simulation as the training set, their prediction power declines they are cross-tested on another simulation. With this simple exercise, we wish to understand whether this deterioration in accuracy is associated with a certain information channel. To this end, we train multiple CNNs with different subsets of the spatially-resolved input maps. To remove completely an input channel, we mask all the corresponding images to zero values. This way, the model learns to ignore this channel as it does not contribute to the feature extraction. We decide to train with the following different combinations of training channels in the 2D spatial maps:

\begin{itemize}
    \item All spatial maps (Mass, Velocity, $\sigma$, Age, Metallicity)
    \item Only mass and kinematic spatial maps (Mass, Velocity, $\sigma$)
    \item Only stellar population spatial maps (Age, Metallicity)
\end{itemize}

For each one of the aforementioned combinations, we train five equivalent models with diverse initialization weights and then ensemble their predictions to quantify their importance in regards to inferring the ex-situ stellar mass fraction. The selection of different combinations derives from attempting to assert the predictive power of groups of properties that either involve mass, kinematic or stellar population information. For every setup, we train the models using the two different training sets to quantify the accuracy of the models when tested and trained on the same as well as across simulations.

\begin{figure*}
    \centering
    \subfloat[Evaluation on the TNG100 test set]{\includegraphics[width=0.9\linewidth]{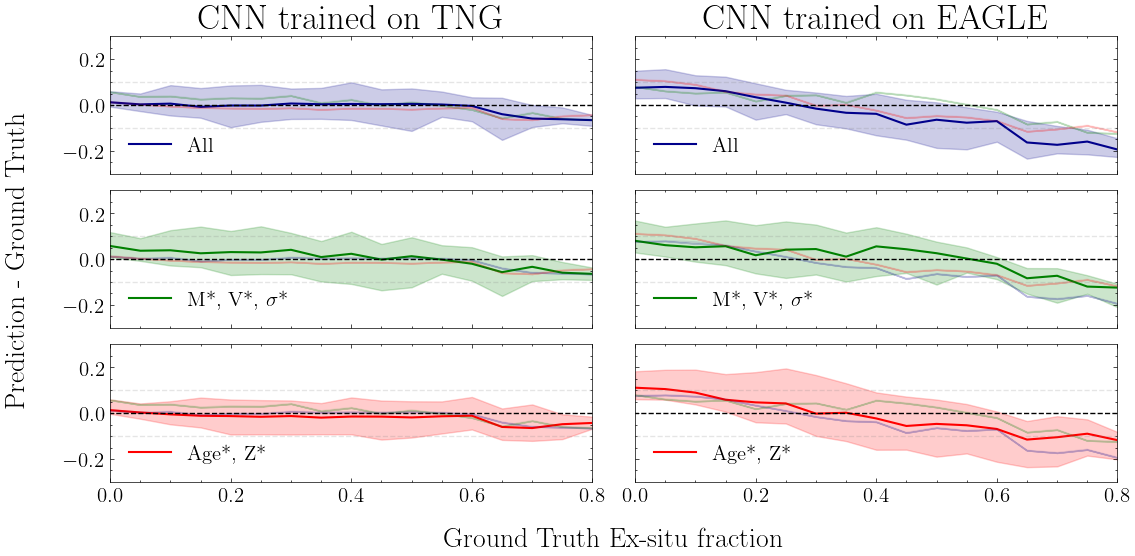}} \\
    
    \vspace{0.3cm}
   
    \subfloat[Evaluation on the EAGLE-L100 test set]{\includegraphics[width=0.9\linewidth]{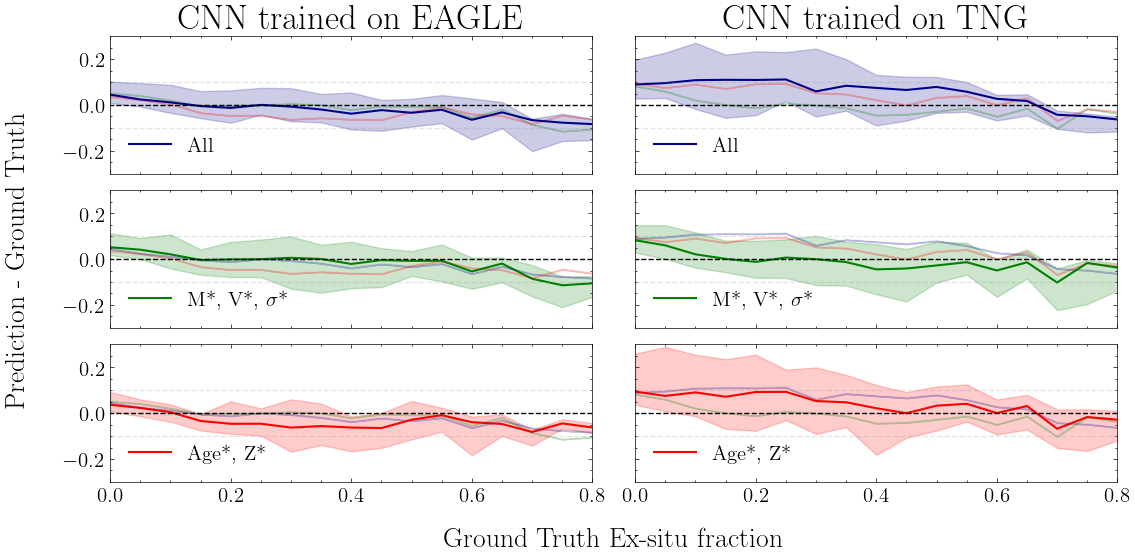}}
   
    \caption{The prediction error (prediction - ground truth) versus the ex-situ stellar mass fraction ground-truth for all galaxies in the TNG100 (a) and EAGLE test set (b) for three different combinations of the input channels. For each panel the median of the over- or under-prediction is illustrated as a solid line and the shaded regions include the 68 per cent of the data points. The medians of the other channels combinations are also displayed in every panel for visual reference. \textbf{Top:} The prediction errors of models trained on three different input combinations when applied on the TNG100 test set. \textbf{Bottom:} The prediction errors of models trained on three different input combinations when applied on the EAGLE-L100 test set. In both sub-figures, we illustrate on the first column the results when testing and training in a single simulation setup and the second column corresponds to the cross-testing scenario across simulations. Each row corresponds to a distinct configuration of the input channels in the 2D maps. Spatially-resolved maps of stellar mass and kinematics prove to be the most robust predictors across simulations.}
    \label{fig:results_relative_error_channels}
\end{figure*}

In Figure \ref{fig:results_relative_error_channels}, we plot the prediction errors, calculated as the difference between the predicted values and the ground-truth, versus the true ex-situ fraction for all galaxies in the TNG100 (a) and the EAGLE-L100 (b) test set. For both figures, the first row displays the results of models trained on all input channels, the second row corresponds to models using only the mass and kinematics information and the third row illustrates the prediction errors of models trained using only the age and metallicity channels. 

First, we evaluate the results on the TNG100 test set. We find that, within a single simulation scenario, the ex-situ stellar mass fraction is consistently predicted with a very low error on all channel combinations when testing and training on TNG100 (Figure \ref{fig:results_relative_error_channels} (a) column 1). Including all channel information renders the lower bias, but this bias remains significantly low (< 0.1) even when training only with the shapes present in a subset of the input channels from the 2D maps. On the lower end and the high-end of the ex-situ fractions, we find that there exists a slight over- and under-prediction respectively, that is less apparent when the stellar population properties are included in the spatial input maps. \looseness-2

Attempting to evaluate the cross-test performance on the TNG100 test set, as illustrated in the second column of Figure \ref{fig:results_relative_error_channels} (a), we observe an increased residual in the predictions of the TNG100 test set from models trained on EAGLE-L100. We find a slight under-prediction in the middle and higher ex-situ values, particularly when age and metallicity are included during training. Notably, models trained only with mass and kinematic information demonstrate a lower bias that is well-bounded between (-0.1, 0.1), as well as a lower scatter than the other channel combinations.

We proceed with the evaluation of the performance of the CNN models on the EAGLE-L100 test set, as illustrated in Figure \ref{fig:results_relative_error_channels} (b). We find that models trained on the same simulation seem to produce a similar behavior in all channel combinations. More specifically, while the prediction errors appear low and well-bounded between the (-0.1, 0.1) space, there exists a consistent under-prediction on the high-end of the ex-situ stellar mass fractions that is evident on all input combinations. A possible reason for that is that the EAGLE-L100 training set is still imbalanced towards high ex-situ fraction galaxies. \looseness-2

Moving to the cross-testing setup, we now find more outstanding differences among the channel combinations upon attempting to cross-test the EAGLE-L100 test set on models trained on TNG100. More specifically, there exists a high over-prediction in all ex-situ regimes from models trained on all input channels as well as from models trained only on age and metallicity (bias > 0.1 for the majority of galaxies with an ex-situ fraction $<0.6$). However, this domain drift appears to be eliminated when only the mass and kinematic information is utilized during training, as the bias is minimized (<0.1) and the scatter is significantly smaller. We pinpoint the stellar population properties as the culprit for reducing the generalisation power of the model. This result suggests that the TNG100 simulation contains populations that have a stellar age and metallicity distribution that resembles galaxies with lower ex-situ stellar fractions in EAGLE-L100. However, their mass and kinematic information is more closely related to galaxies with equivalent merging histories and thus can be used for cross-inference.

Overall, we find that in a fixed simulation, the ex-situ stellar fraction can be predicted with a high accuracy from shapes/gradients present even only in a subset of 2D spatial maps, e.g. only from stellar population properties. Across simulations, though, we find that the spatial distributions of stellar mass and kinematics prove to be more robust predictors, possibly because they contain information more related to gravity.

\subsection{Effect of aperture}

The spatially-resolved input maps have been created with an aperture of 4 stellar half-mass radius ($r_{e}$). Attempting to test the limits of the model, we train a number of CNNs while decreasing the aperture of the input observables. This adjustment is implemented by creating a circular mask of the corresponding radius and masking all pixels outside the mask to zero. We choose to test the performance of the model with the following apertures:
\begin{itemize}
    \item 4 $r_{e}$
    \item 2 $r_{e}$
    \item 1 $r_{e}$
\end{itemize}

For each aperture, we re-train 5 equivalent models, varying only their weight initialization scheme, with the masked training dataset and using all channels. Subsequently, we evaluate the corresponding predictive power by ensembling for each galaxy the inference of these models. It should be noted though that the ground-truth we wish to infer is the ex-situ stellar mass fraction that corresponds to the entire galaxy and not just the inner parts included in the aperture. We wish to understand if there exists sufficient information in the inner parts of the galaxy to predict the global ex-situ stellar mass fraction.

\begin{figure*}
    \centering
    \subfloat[Evaluation on the TNG100 test set]{\includegraphics[width=0.9\linewidth]{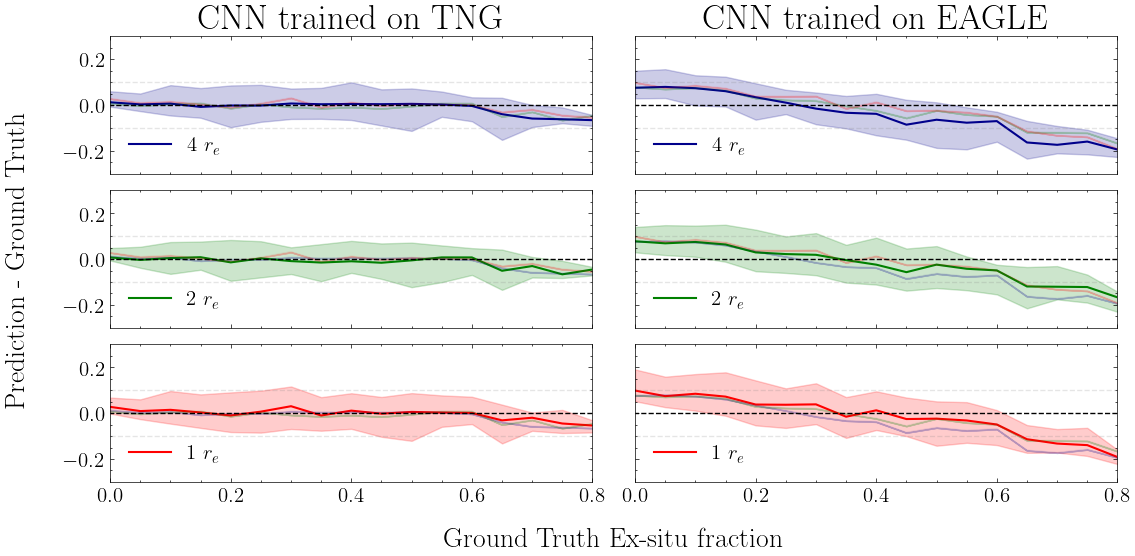}} \\
    \vspace{0.3cm}

    \subfloat[Evaluation on the EAGLE-L100 test set]{\includegraphics[width=0.9\linewidth]{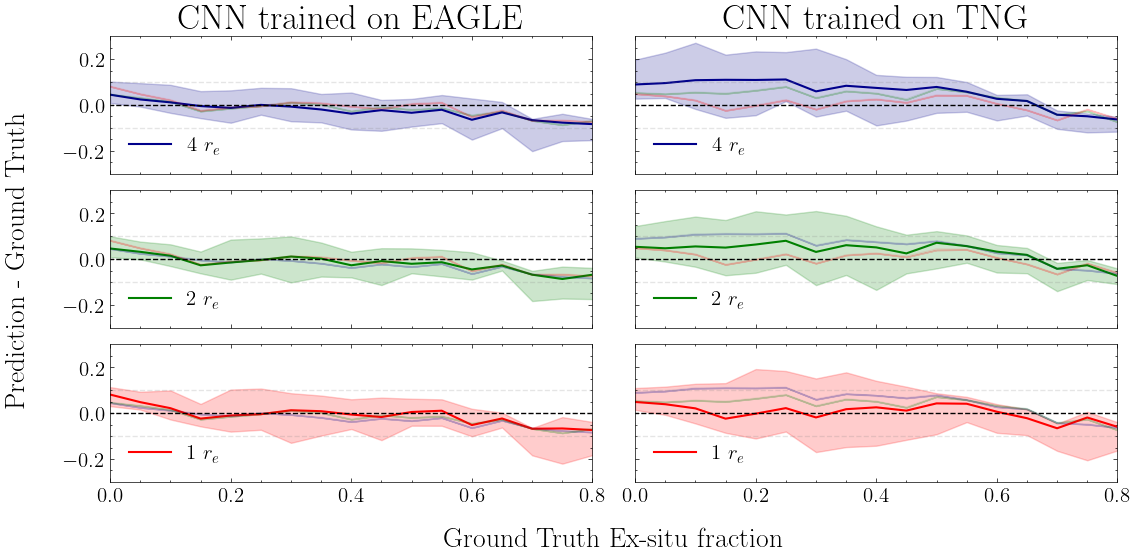}}
    \caption{The prediction error (prediction - ground truth) versus the ex-situ stellar mass fraction ground truth for all galaxies in the TNG100 (a) and EAGLE test set (b) for three representative apertures. The curves and shades are drawn in a similar fashion to Figure \ref{fig:results_relative_error_channels}. \textbf{Top:} The prediction errors of models trained on three different apertures when applied on the TNG100 test set. \textbf{Bottom:} The prediction errors of models trained on three different apertures when applied on the EAGLE-L100 test set. In both sub-figures, we illustrate on the first column the results when testing and training in a single simulation setup and the second column corresponds to the cross-testing scenario across simulations. Decreasing the aperture preserves the predictive power of the models in a fixed simulation while enhancing prediction in the cross-testing.}
    \label{fig:results_relative_error_fov}
\end{figure*}

In Figure \ref{fig:results_relative_error_fov}, we collect the 3 aperture experiments' results for the TNG100 (a) and the EAGLE-L100 test set (b). We first evaluate the aperture effect when training and testing on the same simulation. For the TNG100 test set, we find that the models have a very strong inferring power even when only very central regions of the galaxy are used (e.g. with an aperture of 1 $r_{e}$) with a bias $\sim0$ and scatter $< 0.1$. This suggests that there exists sufficient information in the central regions in the TNG100 simulation for inferring the total ex-situ stellar fraction.

We proceed by evaluating the effect of the changing aperture on the TNG100 test set through the cross-testing scenario (Figure \ref{fig:results_relative_error_fov} (a) second column). We find that the prediction power of the models follows the same trend on all apertures. More specifically, middle and higher ex-situ fractions tend to be slightly under-predicted by the EAGLE-L100 models when tested on the TNG100 test set, especially when information included in 4 $r_{e}$ is utilized. Notably, decreasing the aperture appears to partly reduce the bias in the higher end of the ex-situ fractions.

Moving to the EAGLE-L100 test set, we again find that models trained on the EAGLE training set have a consistent high prediction power independent of the aperture (bias < 0.1 and scatter < 0.15). This suggests that the inner parts of the galaxies in the EAGLE-L100 simulation also contain information relative to the total ex-situ stellar mass fraction that is adequate for an accurate prediction. These consistent results among both simulations in the single simulation setup are very promising for the future application of the models on actual observational data from IFU surveys that usually cover an FOV of $\sim ~ 1.5$ effective radius ($\rm{Re}$) \citep{emsellem_2004, Bundy_2015, FB_2017}. \looseness=-2

Upon evaluating the inferring power of models trained on TNG100 and tested on EAGLE-L100, surprisingly we find that decreasing the aperture reduces the bias of the model by 0.1 - 0.15 for all ex-situ fractions. More specifically, as it was already established on the initial cross-testing experiments, testing EAGLE-L100 on models trained on TNG100 provided a consistent over-prediction of the ex-situ fraction. We find that this domain drift seems to decrease when the models only have access to more inner parts of the galaxies. This indicates that in the outer parts there exist features in TNG100 that are not present in EAGLE-L100 which, while they facilitate a higher precision when tested on the same simulation, do not generalize well across simulations. Taking into account the results from the previous section, it is highly possible that these features can be found mainly in the age and metallicity spatial maps, which proved to be the most problematic on this specific cross-testing scenario.

Overall, we find that reducing the aperture in a fixed simulation setup preserves the high accuracy in the prediction, a result that underlines that one can infer the global ex-situ stellar mass fraction of galaxies using only the gradients present in their inner parts. On the cross-testing scenario, a narrower aperture surprisingly enhances the CNN's robustness. \looseness-2

\subsection{Combined tests}

\begin{figure*}
    \centering
    \subfloat{\includegraphics[width=\linewidth]{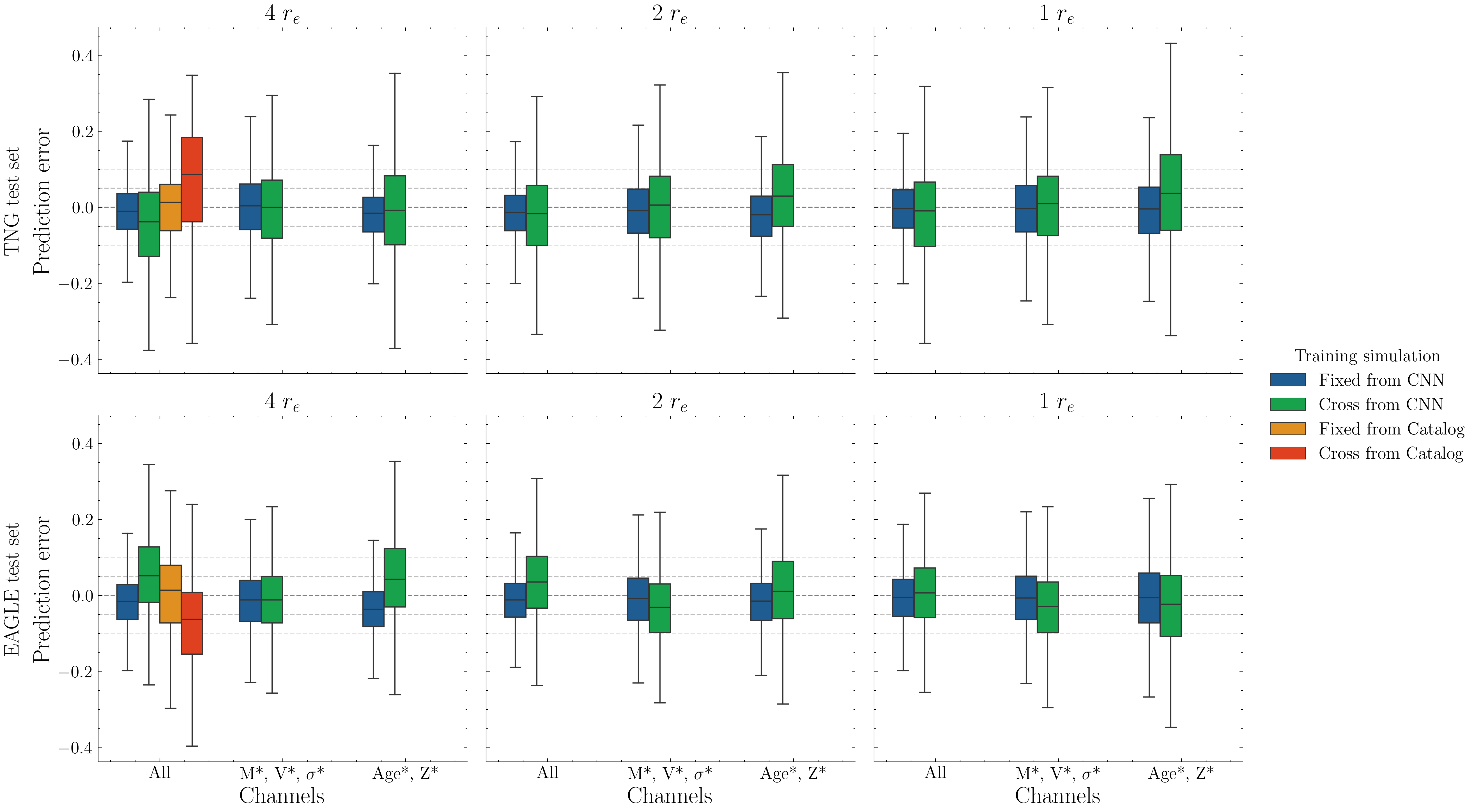}} \\
    \vspace{0.3cm}

    \caption{The distribution of prediction errors between the true values and the inferred ex-situ fractions as they where predicted by the models for all galaxies in the TNG100 and EAGLE-L100 test sets. Each box corresponds to a different training run using the subset of the 2D spatial maps and the aperture annotated. The median of each distribution is indicated with the straight line across the box, while the first and the third quartiles of the distribution correspond to the box limits. The most extreme points of the distribution are illustrated as "whiskers", lines extending from the box. In a fixed simulation, the CNNs provide the best results when all information is available. On the cross-test scenario, the mass and kinematics seem to provide the smaller bias.}
    \label{fig:all_results}
\end{figure*}

For getting a better understanding on the differences emerging among simulations, we run a final set of trainings that integrate the aperture experiment and the different channel inputs combinations. In Figure \ref{fig:all_results}, we compare the prediction error distributions that are produced from each one of the trained CNNs on the TNG100 test set (first row) and the EAGLE test set (second row). We test the effect of aperture on each input channel combination both on the single simulation scenario (blue boxes - Fixed from CNN) as well on the cross-trainings (green boxes - Cross from CNN). For a direct comparison, we also plot the integrated catalog value runs, which are better suited for comparison with the case of the wider aperture and when all input channels in the 2D spatial maps are used (orange boxes - Fixed from Catalog and red boxes - Cross from Catalog). \looseness-2

We first evaluate the effect of the inputs and the aperture on the CNN models trained and tested within a single simulation (blue boxes). We find that the prediction errors present the lowest values when all available information is utilized for training, namely when using the wider aperture (4 $r_{e}$) and all channel inputs for both test sets. This is highly anticipated since the CNN is taking advantage of the additional information and is able to extract the relevant features. All other experiments, though, provide low median prediction error values as well (bias $\sim$ 0 and scatter < 0.1). This suggests that there exist features that are adequate for inference in a single simulation setup even when only very central regions are used from a subset of input channels. Particularly for the TNG100 test set, it appears that the stellar population channels hold substantial information, as their inclusion consistently reduces the scatter of the prediction error.  \looseness-2

We proceed by examining the results of the cross-testing scenarios (green boxes). As it has been discussed above, we now find a significant domain drift across simulations when all information available is utilized (bias increases to $\sim 0.05$ and scatter > 0.15), signifying that the CNN predictive power is affected by the the modelling differences. However, the bias is significantly reduced in both cross-testing setups when only the mass and kinematics information in the 2D spatial maps is used for training (bias $\sim$ 0 and scatter < 0.15 for both cross-tests). With a direct application on IFU observational datasets in mind, we evaluate how the decreasing aperture affects the accuracy across simulation when only the 2D spatial distributions of mass and kinematics are provided as inputs. We find that EAGLE-L100 trainings when tested on TNG100 are not greatly affected by the different apertures or the channel inputs. The scatter seems to grow slightly ($\sim0.15$) when the aperture is decreased or when less information is available from the spatially-resolved input maps. Upon evaluating the EAGLE-L100 test set on models trained on TNG100, the bias and scatter are remaining consistently low when a narrower aperture is used along with the stellar mass and kinematics spatial maps (bias $\sim0$ and scatter <0.15). Overall, we find that decreasing the aperture preserves a high prediction power across simulations when only gradients present in the spatially-resolved mass and kinematics maps are utilized. This result is a very positive indicator that our models will remain robust upon testing on actual observational maps covering an aperture of $\sim$ 1.5 Re. Moreover, since the inference is based only on mass and kinematics gradients, we can remove an extra layer of uncertainty in the derivation of integrated properties of galaxies or the need of applying stellar population models prior to inference. \looseness-2

An additional insight from the combined tests is that the behavior across simulations is not entirely symmetric. As far as the TNG100 test set is concerned, we find that the bias in predictions is not highly affected by the different experiments when models trained on EAGLE-L100 are used for cross-testing. This suggests that the domain drift is not related to a specific channel input or a galactic region, and might be exhibited simply because one simulation contains populations with a particular merging history that are not present in the other. Such a difference can lead to the CNN learning more complex features from the training set and by extension not allowing it to generalize well on unseen domains. The second culprit for the observed domain drift are the age and metallicity gradients/shapes in the outer regions. We argue that the two simulation models produce differences in the shapes observed for galaxies with similar merging histories, more evident when TNG100 is cross-tested on the EAGLE-L100 test set. This indicates that there might exist more prominent gradients in the metallicity and age of galaxies in the TNG100 simulation that the CNN models take advantage of. We will investigate that further in section \ref{gradients}.

\begin{figure*}
    \centering
    \subfloat[Train with $\rm{M}$*, V*, $\sigma$*, Age*, Z* spatial maps within an aperture of 4 $r_{e}$]{\includegraphics[width=0.85\linewidth]{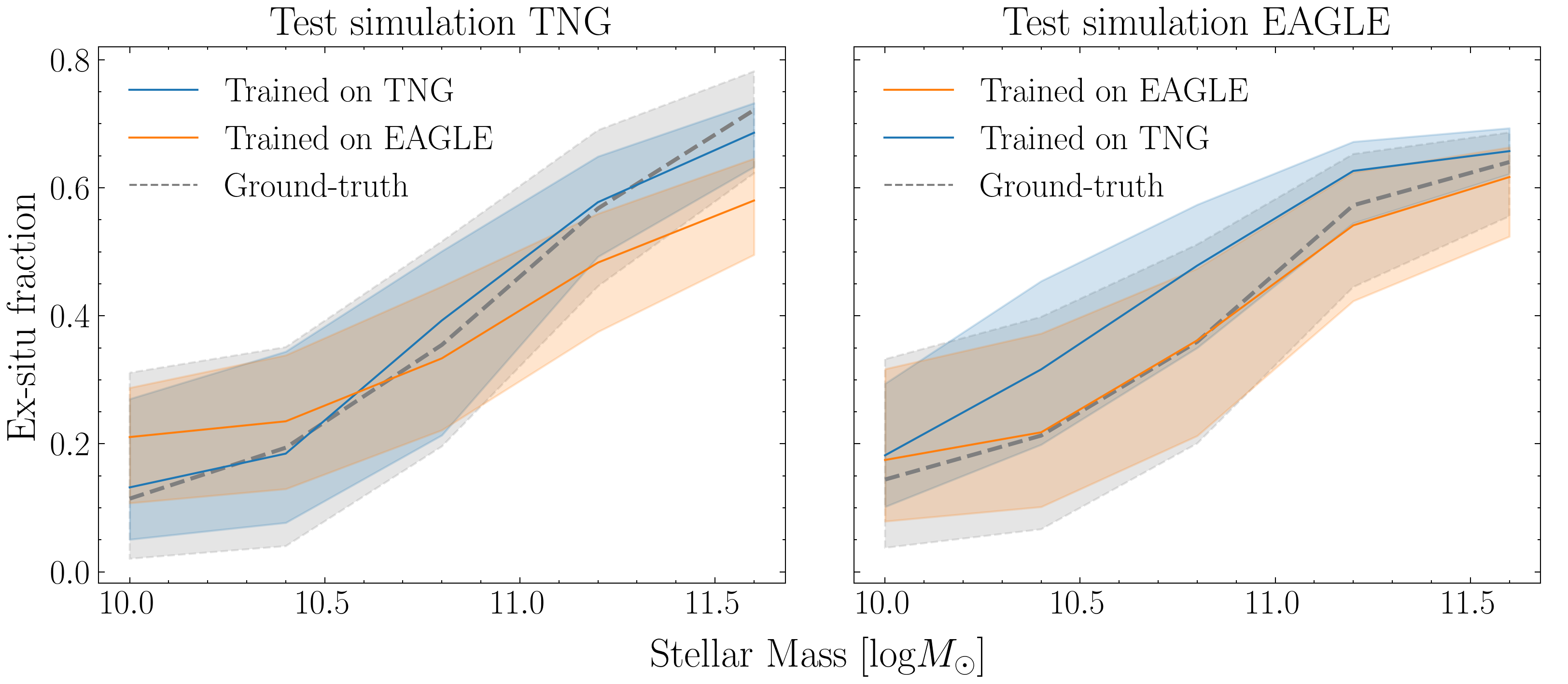}}\\ 
    \subfloat[Train with $\rm{M}$*, V*, $\sigma$* spatial maps within an aperture of 1 $r_{e}$]{\includegraphics[width=0.85\linewidth]{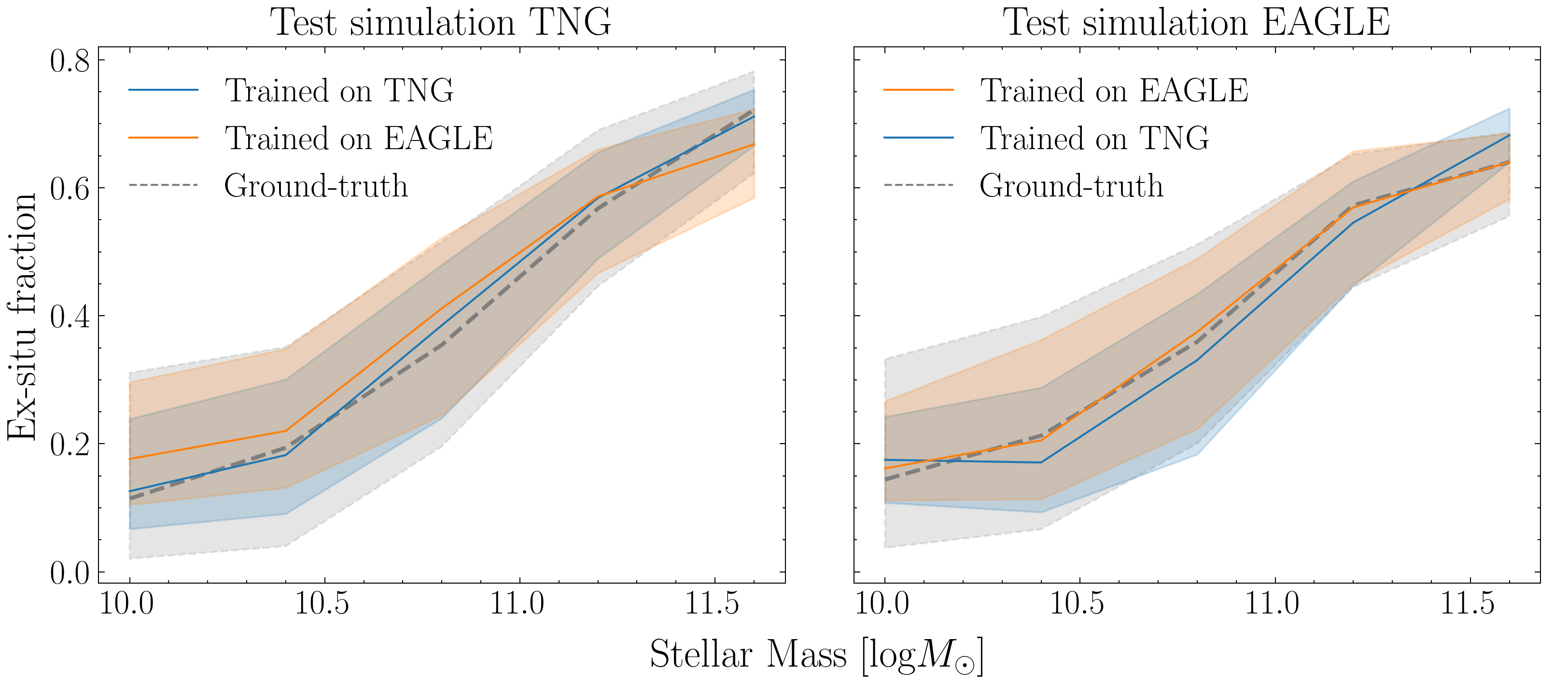}} 
    \caption{The recovery of the ex-situ stellar mass fraction vs. stellar mass relation from the CNN models  on the TNG100 and on EAGLE-L100 test set. The grey dashed line corresponds to the median of the ground-truth ex-situ vs. stellar mass, as it is resolved from all galaxies in the test set of each simulation. The median of the recovery from the predictions of models trained on TNG is drawn with a blue solid line and from models trained on EAGLE with an orange solid line. The shaded regions contain the 68 per cent of all data. \textbf{Top:} The recovery of the stellar mass vs. ex-situ fraction relation from models trained on mass, kinematics, age and metallicity 2D spatial maps within 4 $r_{e}$ aperture. \textbf{Bottom:} The recovery of the stellar mass vs. ex-situ fraction relation from models trained only on mass and kinematics 2D spatial maps within 1 $r_{e}$ aperture. In a fixed simulation scenario there is a fine recovery of the relation for all inputs and apertures. The cross-trained models can not accurately reconstruct the ex-situ vs. stellar mass relation across simulations when stellar populations are included in the inputs in a 4 $r_{e}$ aperture. However, models trained only on mass and kinematics in a 1 $r_{e}$ aperture demonstrate robustness across simulations, with a fine recovery of the ex-situ stellar mass fraction vs. stellar mass relation in a fixed and cross-testing scenario.}
    \label{fig:mass_recovery}
\end{figure*}

\section{Towards a robust model across simulations} \label{sec:robust_model}

So far, we have explored the effect of aperture as well as of different input combinations on the accuracy of models trained to predict the ex-situ stellar mass fraction across the TNG100 and the EAGLE-L100 simulation. We have found that including only mass and kinematics 2D spatial maps in a reduced aperture (1 $r_{e}$) provides the most unbiased predictions across the two cosmological simulations. We wish to further investigate the robustness of these models, so that we can determine if they are suited for application on actual observational data.

\subsection{Ex-situ vs. Stellar Mass recovery} \label{recovery_original}

A first crucial evaluation metric of the models' performance is how well they can recover the known relationship between the ex-situ stellar mass fraction and the stellar mass. This relation is not identical across the two cosmological simulations in focus. Thus, it is desirable that a trained model is able to reconstruct it irrespective of the origin of the input data utilized during training.

In Figure \ref{fig:mass_recovery}, we plot the recovery of the stellar mass vs. the ex-situ stellar mass fraction relation as it was resolved from the predictions of the trained models when applied on the TNG100 and the EAGLE-L100 test sets. To evaluate better the improvement from the best model configuration across simulations (training only on mass and kinematics within 1 $r_{e}$ aperture), we illustrate the respective results for the original training setup (training on mass, kinematics, age and metallicity inputs within 4 $r_{e}$ aperture) in the upper row of Figure \ref{fig:mass_recovery}. The corresponding results of the best trained models can be found in the bottom row. For a direct comparison with the ground-truth from the cosmological simulations, we plot in all panels the median of the relation between the actual ex-situ stellar fraction vs. stellar mass for all galaxies in the corresponding test set with a dashed grey line. \looseness-2

In the original training setup, we observe that in a fixed simulation scenario the predictions trace very closely the ground-truth vs. stellar mass relation. This reconfirms the prediction power of the normalized spatially-resolved maps in a single cosmological setup. However, we find that training with stellar mass, kinematics, age and metallicity does not guarantee a fine recovery of the ex-situ stellar mass relation across simulations. More specifically, models trained on EAGLE under-predict all ex-situ fractions for stellar masses $> {10}^{10.5}\ {M}_{\odot }$ by almost 20 per cent, while over-predicting the accreted mass for all galaxies with stellar mass $< {10}^{10.5}\ {M}_{\odot }$. Additionally, models trained on TNG consistently over-predict the ex-situ stellar mass fraction of EAGLE galaxies across all stellar masses by 10-15 per cent. 

On the contrary, models trained only on mass and kinematic gradients within a 1 $r_{e}$ aperture (Figure \ref{fig:mass_recovery} (b)) demonstrate a powerful recovery of the ex-situ fraction vs. stellar mass relation on both tests, independent of the training set used during training. Even the cross-trained models trace the ground-truth curve very closely for all galaxies across the considered stellar mass range. This result further underlines that the trained models are robust across simulations and provides a strong indication that such models are suitable for application on actual observational data.

\subsection{Uncertainty calibration} \label{uncertainty}

A fundamental aspect that should always be thoroughly considered prior to applying simulation-based inference on real data is how one can provide a measure of uncertainty on the resulting predictions. Therefore, throughout this work we employ probabilistic neural networks that allow an evaluation of the confidence on the inference. However, it should be investigated whether the standard deviation of each posterior can serve as a meaningful measure for capturing the uncertainty of the model. 

We need to ensure that the learnt posterior is calibrated properly by producing relatively conservative results. To do so, we assess the coverage probabilities through a statistical coverage test \citep{coverage}. In practise, we explore whether the predicted posteriors are accurate by investigating what percentage of times the error is within the x\% of the posterior width. For an accurate posterior, this percentage should be around x\% (e.g. 68 per cent of galaxies from the test set should have a prediction error $\leq 1 \sigma$, corresponding to a 68 per cent of the probability volume).

We compute the empirical coverage over the set of samples, defined as the fraction of samples whose true parameter values fall within the confidence interval we examine at the time. We evaluate the confidence for the 5 equivalent models so that the statistically uncertainty of the results is also quantified. Since we mainly target for a domain-invariant inference using two cosmological simulations, we repeat the calibration test for each trained model both on the TNG100 and the EAGLE-L100 test sets.

\begin{figure}
    \centering
    \subfloat[Train with $\rm{M}$*, V*, $\sigma$*, Age*, Z* spatial maps within an aperture of 4 $r_{e}$]{\includegraphics[width=\linewidth]{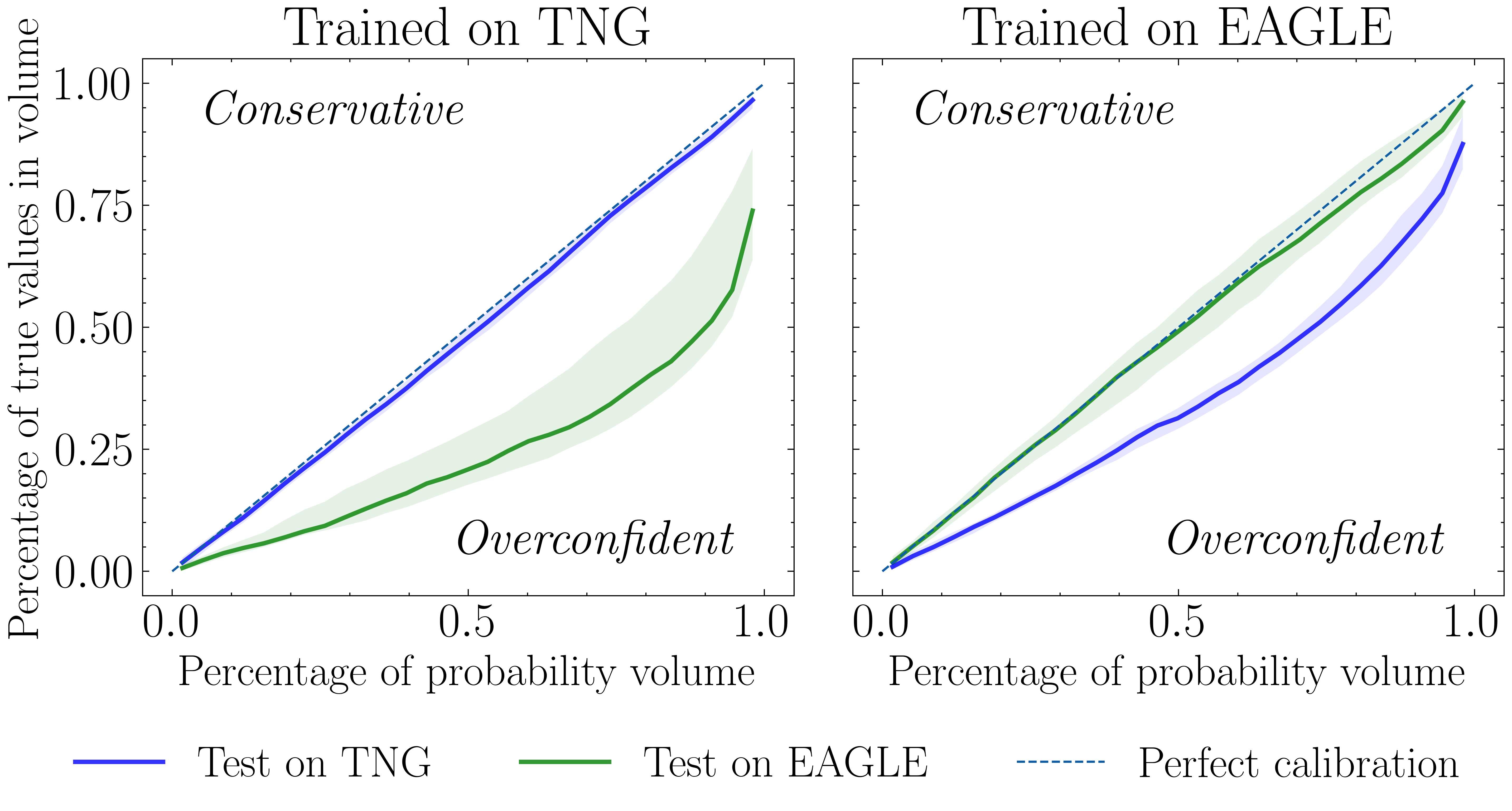}}
    \\
    \subfloat[Train with $\rm{M}$*, V*, $\sigma$* spatial maps within an aperture of 1 $r_{e}$]{\includegraphics[width=\linewidth]{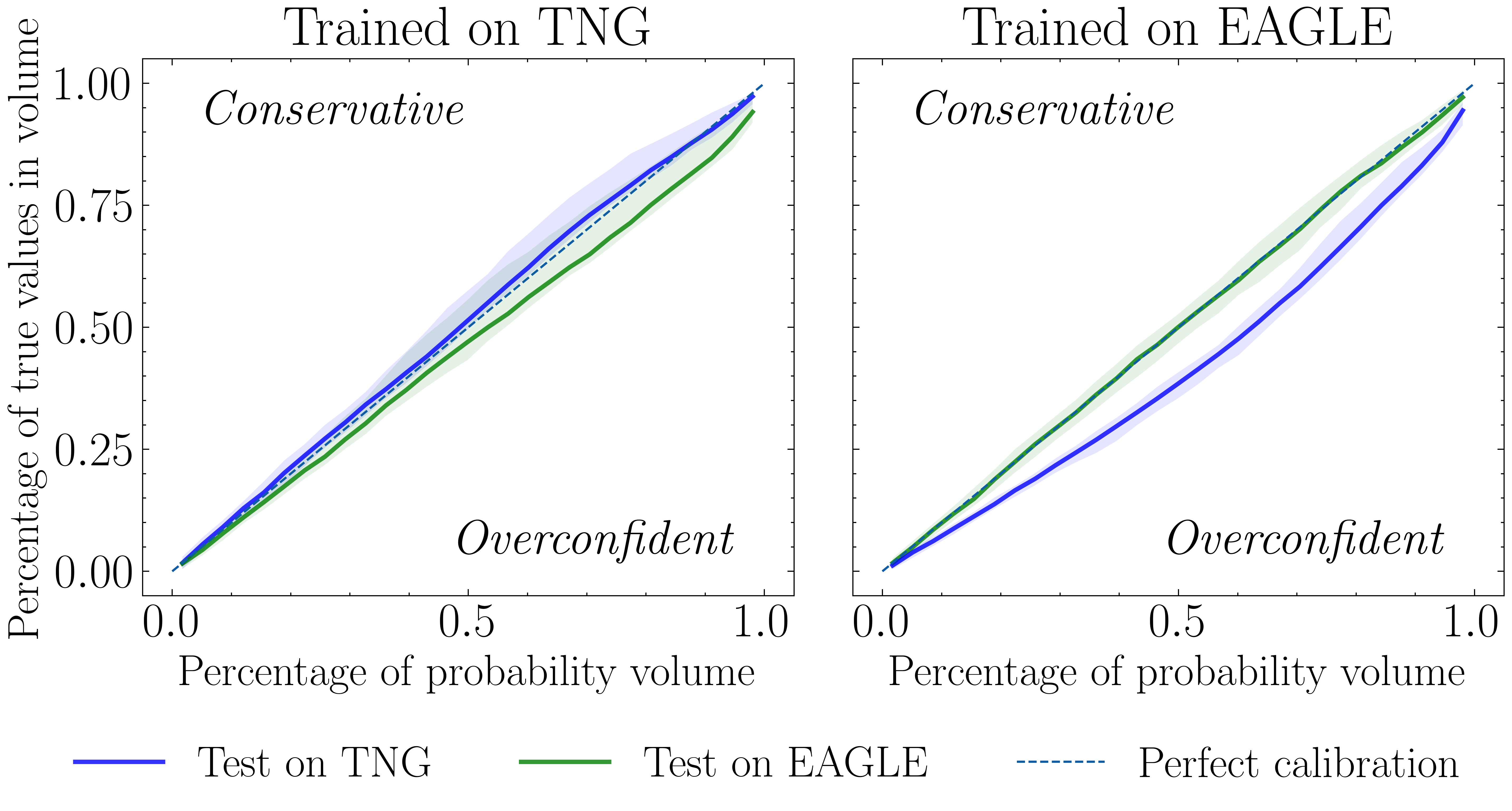}}
    
    \caption{The probability coverage of the CNN models when trained on TNG100 and on EAGLE-L100 for both test sets. For each trained model, we calculate the empirical coverage on all galaxies from the respective test set. The solid blue line corresponds to the median of the uncertainty calibration of the TNG test set and the green to the EAGLE test set. The perfect calibration is drawn with a diagonal dashed line. Models should trace closely the perfect calibration to provide a meaningful measurement of uncertainty, otherwise they are either conservative or overconfident.   
    \textbf{Top:} The calibration tests from models trained on mass, kinematics, age and metallicity 2D spatial maps within 4 $r_{e}$ aperture. \textbf{Bottom:} The respective uncertainty tests for models trained only on mass and kinematics 2D spatial maps within 1 $r_{e}$ aperture. On a single simulation scenario, both estimators closely trace the perfect calibration regime, while having the desired property of being slightly conservative. Across simulations, the estimators originating from the original models appear highly overconfident. On the contrary, when only mass and kinematics in a narrower aperture are employed, the models provide a probability coverage significantly closer to the perfect calibration on the cross-testing. }
    \label{fig:coverage_cross}
\end{figure}

In Figure \ref{fig:coverage_cross}, the coverage test results are illustrated. For a better comparison, we plot the respective results for the original trained models on mass, kinematics, age and metallicity in 4 $r_{e}$ aperture (a) and the best models across simulations trained only on spatially-resolved mass and kinematics maps in 1 $r_{e}$ aperture (b). We run the coverage test for both test sets and the medians of the results are illustrated for TNG100 with a blue solid line and for EAGLE-L100 with a green solid line. Additionally, the shaded regions correspond to the middle-68 per cent containment among the different 5 models. The perfect calibration is illustrated with a dashed blue line.

For the original training setup, we find that testing on a dataset originating from the same simulation as the training set produces an empirical coverage that is very close to the perfect calibration. Moreover, especially for TNG100, the test coverage reveals that the models are slightly conservative, which is a desired property. However, upon applying the coverage test on a cross-test dataset it is apparent that the posteriors are highly overconfident. This is especially evident on the scenario of evaluating a TNG100 trained model on an EAGLE test set and it agrees with the lower accuracy results we found in section \ref{ground_truth_cross}. It is also especially worrying, as it confirms that the bias observed in the model when changing domains is not translated into higher uncertainties.

Contrarily, we find a fine uncertainty calibration from the best models across simulations for almost all considered cases as shown in Figure \ref{fig:coverage_cross} (b). Similarly to the previous setup, in a fixed simulation scenario the empirical coverage traces very closely the perfect calibration. More importantly though, we find that the standard deviations inferred from cross-trained models now demonstrate a behaviour that better resembles a reliable measure of uncertainty on the predictions. Models trained on TNG100 exhibit an almost perfect probability coverage when applied on the EAGLE-L100 test set. This result is particularly encouraging, especially when compared with the high overconfidence that was demonstrated from the original models (left panel of Figure \ref{fig:coverage_cross}(a)). Additionally, models trained on EAGLE are significantly closer to the perfect calibration when tested on TNG100, while they remain still slightly overconfident. 

Overall, we find that CNN models trained on mass and kinematics gradients within an aperture of 1 $r_{e}$ not only demonstrate a higher prediction power across simulations, but also prove to be more robust in their uncertainty measurements. This is crucial for a potential application on actual observational data, as every prediction should be accompanied with a meaningful error estimation. \looseness-2

\section{Discussion} \label{sec:discussion}

\subsection{What do the models learn?}

\subsubsection{Representation space for both simulations}
While the use of neural networks allows the extraction of more complete information, interpretability of what the models learn is not directly available, as they quite often work as black-boxes for the user. Nonetheless, we can still gain a better insight on what the models extract from the 2D spatial maps by exploring the distribution of representations in the CNN layers. \looseness-2 

We utilize the last convolution layer of our trained CNNs to map all galaxy objects from the test set to a high-dimensional feature space that is representative of the feature extraction the model was trained on. We then use the Uniform Manifold Approximation and Projection for Dimension Reduction \citep{umap} (UMAP) to embed this space into a better-suited for visualization 2D-space and color-code it with the ground-truth value of the property we wish to infer, namely the ex-situ stellar mass fraction of each galaxy. This allows a visual inspection of whether the network is able to learn a representation space that is generally smooth in terms of the ex-situ fraction. Additionally, UMAP allows us to better understand why there exists the observed domain drift upon moving from one simulation to the other. More specifically, we can visualize how the representation space behaves on the cross-testing scenarios across simulations with a direct comparison between the two spaces.

\begin{figure}
     \subfloat[UMAPs of last convolution layer of model trained on TNG]{\includegraphics[width=\linewidth]{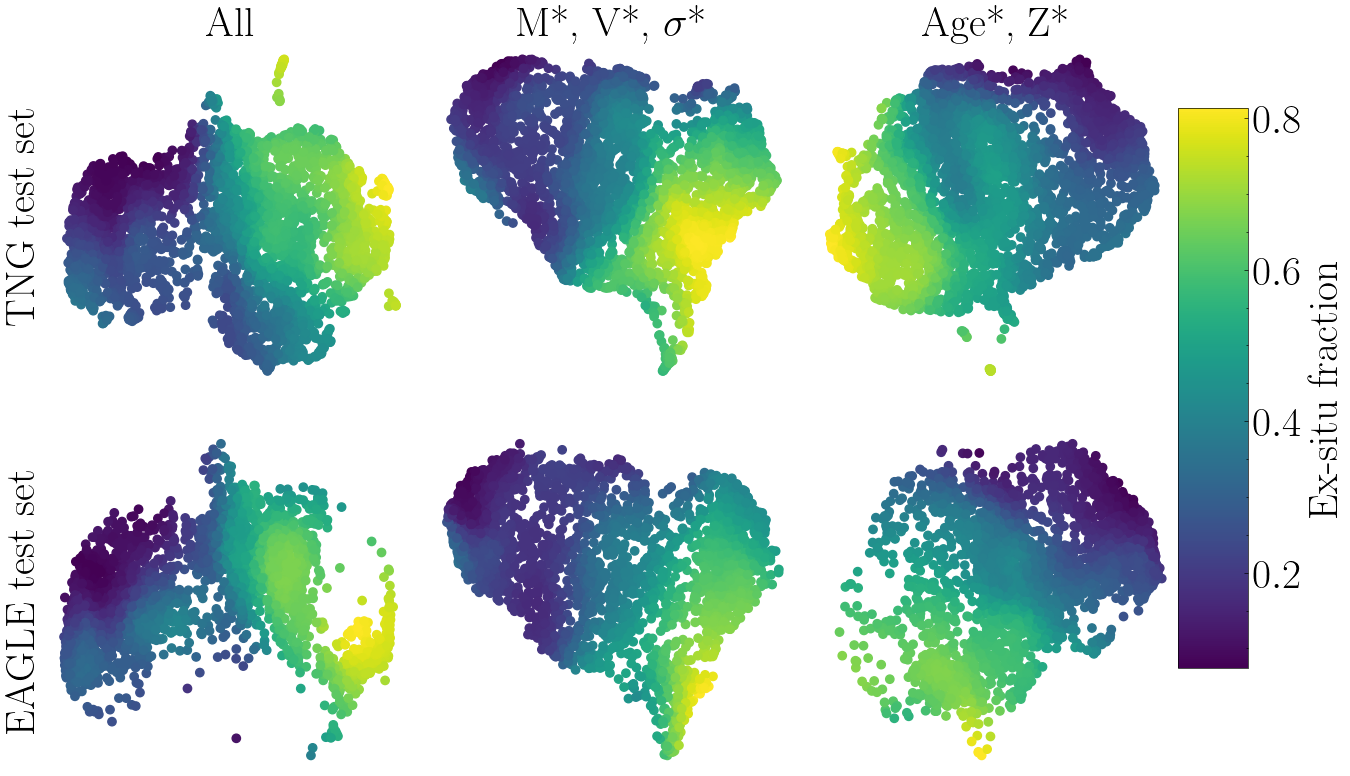}} \\
     \subfloat[UMAPs of last convolution layer of model trained on EAGLE]{\includegraphics[width=\linewidth]{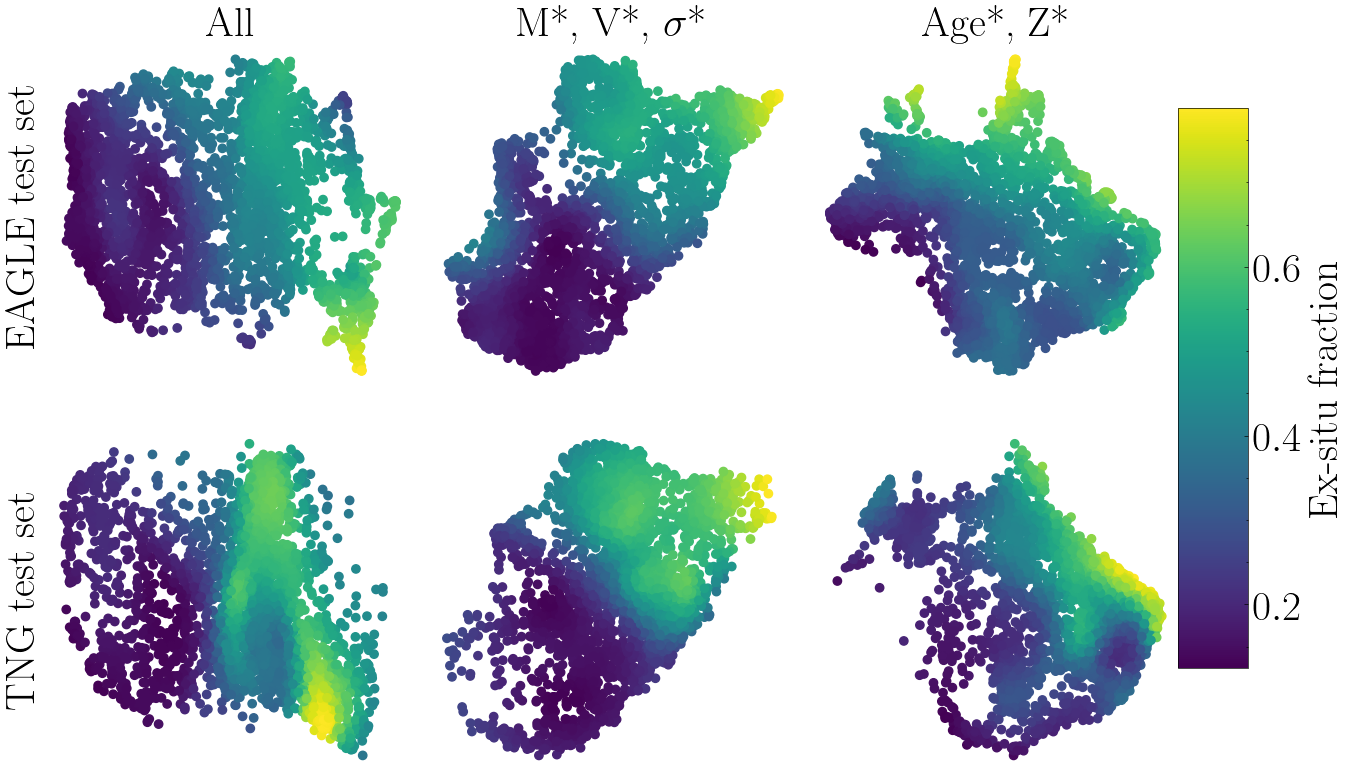}} \\
    \caption{UMAPs of the last convolution layer of CNN models trained on TNG100 (a) and EAGLE-L100 (b) using a wide aperture (4 $r_{e}$) on the training 2D spatial maps when applied on the corresponding test sets. For each training set, we variate the input channels and produce the UMAPs for all galaxies present in the TNG100 and EAGLE-L100 test set. We color-code the embeddings with the true ex-situ stellar mass fraction of the galaxy that each point corresponds to in the representation space. \textbf{Top:} The UMAPs embeddings as they were resolved from the outputs of the last convolution layer of models trained on TNG when applied on the two test sets. \textbf{Bottom:} The UMAPs embeddings as they were resolved from the outputs of the last convolution layer of models trained on EAGLE-L100 when applied on the two test sets.} 
    \label{fig:umaps}
\end{figure}

In Figure \ref{fig:umaps}, the obtained UMAPs are shown for models trained on TNG100 (a) and EAGLE-L100 (b). We decide to apply the UMAPs on three different models for each training set, one trained with all available input channels (first column), one trained only with stellar mass and kinematics information (second column) and one using only the age and metallicity channels (third column). Since the channels experiment revealed a high domain drift among simulations, we wish to understand if this is also reflected on the feature extraction. For each model, we illustrate the representation space both within a single simulation scenario (first row) and the cross-test setup (second row).  \looseness-2

We initially evaluate how the UMAPs are behaving when trained and tested on the same simulation (first rows of subfigures in Figure \ref{fig:umaps}). We find that the space is very smooth in regards to the ex-situ stellar mass fraction in all channel combinations, while not being completely linear. This suggests that the networks can successfully extract complex features from the gradients present in the spatially-resolved input maps -- or even a subset of them -- that are sufficient for predicting the ex-situ stellar mass fraction.

Moving on to the cross-testing scheme, we initially compare the two rows of Figure \ref{fig:umaps} (a), corresponding to how a model trained on TNG represents the space on the TNG and EAGLE test set respectively. We find that the representation spaces are very similar and the ex-situ fraction mappings are generally comparable. However, we observe that there exist gaps on the cross-testing spaces, corresponding to the EAGLE embeddings, that are not present when plotting the TNG embeddings. These gaps are more prominent when the spatial maps contain the stellar population properties (first and third column). Contrarily, the model trained only on the mass and kinematic gradients produces very similar embeddings on both test sets, particularly in the low ex-situ regime.

Respectively, we proceed by comparing the representation spaces on the EAGLE and the TNG test sets on models trained on EAGLE (Figure \ref{fig:umaps} (b)). We again find high similarities among the different test set representations. The cross-testing scenario still creates some gaps in the embedded space which -- while less prominent -- exist in all input combinations.

The existence of regions in the representation space that are not covered with a similar intensity when the two test sets are resolved signifies that the trained model has learnt a particular mapping between the inputs and the ex-situ stellar mass fraction that is not present in both simulations. When models are trained on TNG100, such features are clearly originating from the stellar population spatial maps, as the representation space when only the mass and kinematics channels are employed is smooth for both test sets.  On the contrary, the UMAPs from the last convolution layer of models trained on EAGLE-L100 do not provide so straightforward insights. The gaps in the embedding in all channel combinations might suggest that there exist galaxies with a particular merging history that are not shared across simulations.

\subsubsection{Inspecting the gradients} \label{gradients}

\begin{figure*}
    \subfloat{\includegraphics[width=\linewidth]{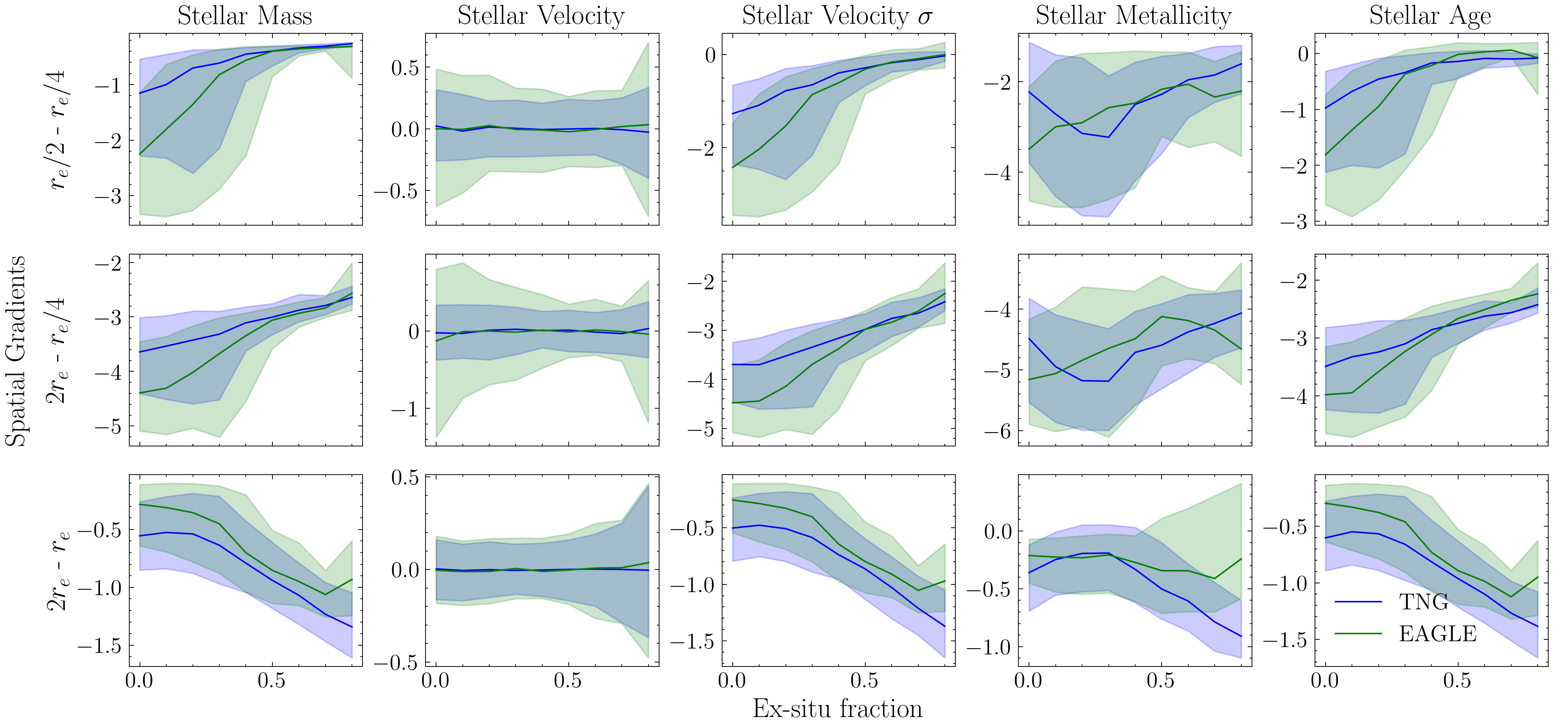}}
 
    \caption{Spatial gradients of the input 2D spatially-resolved maps for three combinations of different radii from the galaxy centre vs the ex-situ stellar mass fraction for all galaxies in the training set of the TNG100 and the EAGLE-L100 simulations. The solid blue and green lines correspond to the median values of all gradients from the TNG and EAGLE simulations respectively. The shaded regions include the 68\% of all datapoints. The gradients between two radii are calculated as the difference between the mean values of all pixels contained in a ring around the two radii on focus (extracted from the standardized spatial maps). While there is a good agreement for the mass, velocity, velocity dispersion and age gradients, we find outstanding differences on the metallicity gradients between TNG100 and EAGLE-L100. This divergence may be the culprit for the domain drift when the stellar population spatial maps are included in the training inputs.} 
    \label{fig:gradients}
\end{figure*}

Throughout this work, we utilize 2D spatially-resolved input maps that have been normalized per image to infer the ex-situ stellar mass fraction. This implies that the information extracted from the CNNs is only present in the shapes or gradients contained in the spatial maps, since all relative value differences per galaxy are removed through the normalization preprocess. In an attempt to investigate further why there exists a domain drift upon testing across simulations when all information is available in the inputs (namely when using mass, kinematics, age and metallicity), we decide to visualize the gradients between the inner parts and outer parts of galaxies on all input channels.  \looseness-2

In Figure \ref{fig:gradients}, we illustrate the spatially-resolved gradients between inner parts ($r_{e}/2$ - $r_{e}/4$), inner versus outer parts ($2r_{e}$ - $r_{e}/4$) and outer parts ($2r_{e}$ - $r_{e}$) for all galaxies from the training sets of both simulations and plot them versus their ex-situ stellar mass fraction. To measure the gradients between two different radii, the mean of all pixel values present in a ring around each radius is calculated and the corresponding gradients are computed as the difference between the two values. The spatial gradients are shown for all five input maps, namely the stellar mass, velocity, velocity dispersion, metallicity and age channels. We note here that the gradients are extracted from the standardized maps, so the corresponding values actually reflect the differences measured in relation to the standard deviation $\sigma$ of each individual image and not the real values.

Firstly, we find that there exists a fair amount of information in the gradients that correlates with the ex-situ stellar mass fraction. This confirms that the CNN most probably uses the gradient features to predict, apart from the velocity channels where the gradients appear flat in all radii. 

Additionally, we find the same trends in the mass, velocity $\sigma$ and age channels for both simulations. More specifically, higher ex-situ galaxies appear to have flatter gradients in the inner regions and inner versus outer regions than lower ex-situ fraction galaxies in both simulations. In the outer regions ($2r_{e}$ - $r_{e}$) the trend is reversed in all of the mass, velocity $\sigma$ and age channels and for both simulations. In these regions, galaxies with lower ex-situ mass fractions have flatter gradients that get steeper in higher ex-situ galaxies. In the respective channels, we also find that the EAGLE simulation has more a more prominent difference between low and high ex-situ fractions and generally a larger scatter. \looseness-2

As far as the metallicity channel is concerned, we find significant differences in the gradients between the TNG and the EAGLE simulation. In the inner regions, TNG exhibits flatter gradients on very low ex-situ galaxies that get steeper up to a point ($\sim0.3$ ex-situ) and then start getting flatter again. Contrarily, on EAGLE the correlation with ex-situ appears to be reversed, as lower ex-situ galaxies have steeper gradients that get flatter towards higher ex-situ galaxies. Very prominent differences appear in the outer regions, where EAGLE produces galaxies with generally flat gradients for all ex-situ fractions, while on TNG higher ex-situ galaxies have clearly steeper gradients.  \looseness-2

These insights agree with the prediction results we obtained from the trained CNNs. We pinpoint the different trends existing in the metallicity gradients between the two simulations as the main origin for the domain drift observed on the cross-testing. Additionally, we argue that CNNs trained on TNG including the metallicity channels with a larger aperure fail to predict efficiently on EAGLE because the network is using gradient information from the training set that is not present in the EAGLE dataset. Also, we find that EAGLE has generally a larger scatter, which might explain why it generalizes better when all channels are utilized and the aperture is around $4 r_{e}$.

\subsection{Comparison with other works}
\begin{figure*}
    \subfloat{\includegraphics[width=\linewidth]{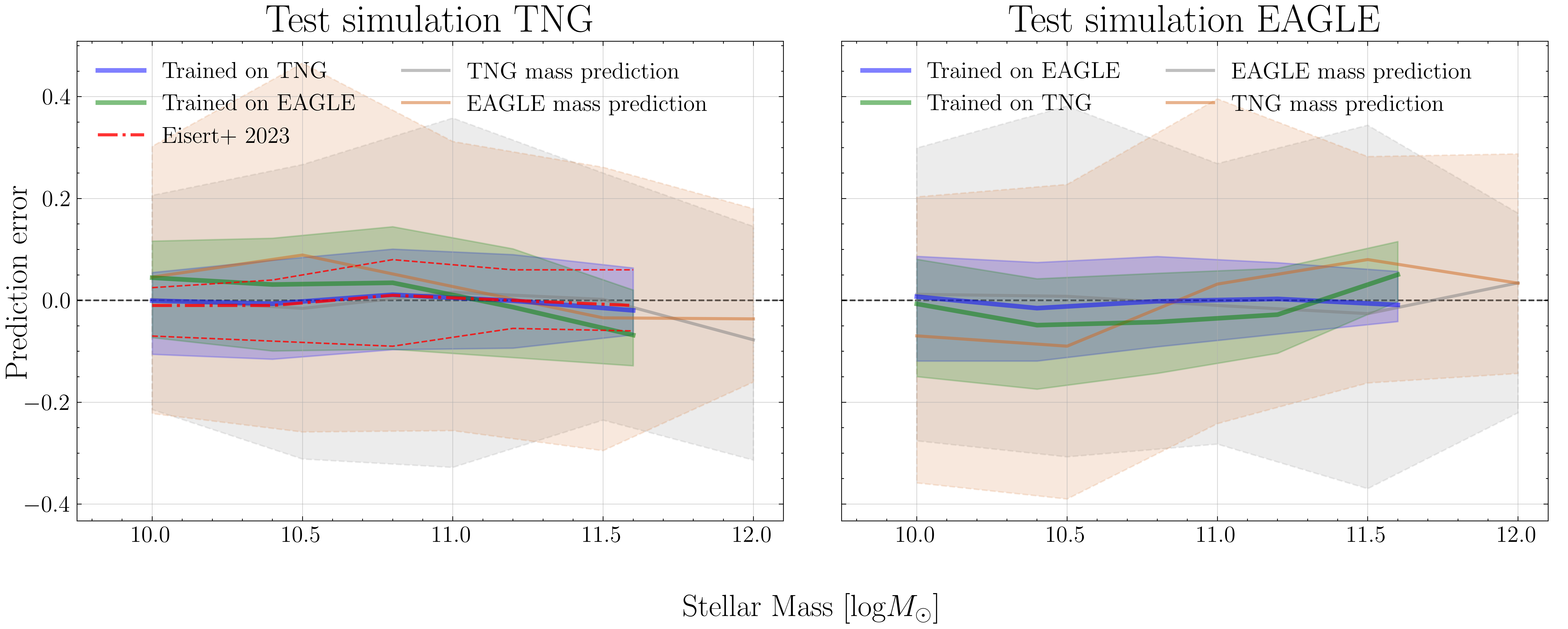}}

    \caption{The prediction accuracy of the models as a function of galaxy stellar mass for the TNG100 and EAGLE-L100 test set. We plot the prediction error (prediction - ground truth) as it is resolved from the predictions of the CNN models trained on mass and kinematic gradients contained within 1 $r_{e}$ aperture for all galaxies in the TNG test set (left) and EAGLE test set (right). The blue and green solid lines correspond to the median error resulting from models trained on the same simulation and cross-trained respectively. The shaded regions contain the $68\%$ of all data. To compare, the median error from \protect\cite{Eisert_2022} is drawn with a red solid line while the $68\%$ containment of all data is marked between the area in the red dashed lines. Finally, we plot the residuals produced when the bare relation between the ex-situ fraction and the stellar mass is employed for predicting the accreted fraction through a simple cubic interpolation for both training schemes as the grey and light brown solid lines (median) and shaded areas ($68\%$ of all data) respectively. The ex-situ stellar mass fraction can be recovered in both simulations with <15\% scatter across the stellar mass range independent of the origin of the dataset used for training.} 
    \label{fig:compare_other}
\end{figure*}   

The main purpose of this work is to investigate whether an indicative quantity of the merging history of galaxies -- the ex-situ stellar mass fraction -- can be inferred accurately across cosmological simulations solely from observable quantities irrespective of the underling sub-grid differences. While we find that using only mass and kinematic information as input can maintain a strong prediction power across the sub-grid modelling variations within the context of this work, it is also important to compare these results with previous works for a more quantitative evaluation.

Closely related to the present paper, \cite{Eisert_2022} utilize machine learning for the inference of various properties describing the merging history of galaxies from observable summary statistics in the TNG100 cosmological simulation. Through conditional invertible neural networks, they manage to obtain predictions for the mean merger time, mean merger mass ratio, last major merger mass and time, along with the ex-situ stellar mass fraction of galaxies. By utilizing all discrete snapshots between $z = 0$ and $z = 1$ and a stricter mass range ($10^{10-12} M_{\odot}$), they employ integrated properties (e.g. stellar mass and redshift)  as inputs to create a massive dataset of $182625$ galaxies. In the present work, along with a set of summary statistics, we also attempt to infer the ex-situ stellar mass fraction of galaxies from 2D observable spatial maps. This approach allows us to build a sufficient training set from a limited amount of galaxies by using multiple realisations per galaxy through distinct 2D projections. Additionally, we differentiate further by removing the mass and size dependencies from the 2D maps and only focus on the spatially-resolved gradients present.  \looseness-2



In Figure \ref{fig:compare_other}, we plot the prediction error as a function of the considered stellar mass range for the CNN models presented in this work. We decide to show the results obtained from models using only the mass and kinematic gradients contained within 1 $r_{e}$ aperture, as these proved to be the most robust predictors across simulations. We compare the residuals from the single simulation training (solid blue lines) and the cross-training (solid green lines) and find that the bias of the residuals is low and comparable across simulations, indicating that the proposed methodology can infer with high accuracy using domain-invariant features. This result suggests that the trained models can be potentially applied on actual observational data with inconsiderable loss in precision. Additionally, we plot the median of the error as it was reported by \cite{Eisert_2022} with a red dot-dashed line and note that a similar bias with the present work is demonstrated, while a slightly lower scatter is reported (dotted red lines). A similar scatter is acquired in a fixed simulation when a wider aperture is utilized (4 $r_{e}$). However, we prefer to sacrifice this extra precision to ensure robustness across the different cosmological models, which is more pronounced with a narrower aperture. We emphasize here that the CNNs demonstrate a comparable prediction power with the integrated approach followed by \cite{Eisert_2022}, while a significantly lower number of galaxies is utilized for training (by a factor greater than 20). 
In a closely-related work, \cite{Shi_2022} use a more comparable training set of central galaxies at $z=0$ from the TNG100 simulation to infer the ex-situ stellar mass fraction though a random forest approach and report a higher bias (up to 5 per cent) and a scatter up to 15 per cent when only observable features are utilized. This further underlines the magnitude of information the spatially-resolved maps hold for inferring the ex-situ stellar mass fraction of galaxies. \looseness-2

Lastly, as noted previously, stellar mass shows a significant correlation with the ex-situ stellar mass fraction for both cosmological simulations in focus. In an attempt to evaluate how accurately one can predict the ex-situ stellar mass fraction solely from that bare relation and measure how our models improve upon that, we use a cubic spline interpolation \citep{Boor1978APG} on all data points from the training sets on the ex-situ vs. stellar mass plane for both simulations. By applying the resulting interpolations to the test sets, we can acquire a measurement of error between the real value of ex-situ and the value predicted using only the stellar mass. Additionally, we can also compute the returned mass predictions across simulations, since the mass vs. ex-situ relation differs between the two cosmological models. We plot the median of the error produced from the resulting interpolation in a fixed simulation as a solid light grey line and the corresponding error from the cross relation as a solid light brown line in Figure \ref{fig:compare_other}. It is apparent that the trained neural network models greatly improve upon this bare relation, with a significantly lower bias and scatter. Notably, the cross-testing results from the CNNs also provide a superior prediction that surpasses by far the ex-situ vs. mass relation even in a fixed simulation scenario and is more comparable to the best case training within a single cosmological model. \looseness-2

\subsection{Using the public version of the EAGLE-L100 data}

The main objective of this work is to examine the impact of different galaxy-formation models on the inference of the ex-situ mass fraction of galaxies from observable properties. To ensure that only the effect of the different sub-grid physics imposed by each cosmological simulation is captured throughout our analysis, we have utilized so far datasets from TNG100 and EAGLE-L100 that have been analysed with identical halo/subhalo and merger tree definitions and only differ in the underlying sub-grid recipes that govern the galaxy-formation models. 

However, it is also worth exploring how the robustness of trained neural network models is impacted across simulations when the training datasets originate from the public versions of the TNG100 and EAGLE-L100 simulations. For that, we need to use the original EAGLE-L100 simulation data that is available from the EAGLE public release along with the (already-used) public dataset from TNG100. This agnostic approach can provide an insight on how the distinct merger tree and subhalo definitions between simulations can affect predictions across simulations, on-top of the underling sub-grid physics.

Both the TNG100 and the EAGLE-L100 public versions identify halos with the friends-of-friends (FoF) algorithm \citep{1985ApJ...292..371D}. Additionally, in both cosmological simulations, subhalos are identified with the Subfind algorithm \citep{2001MNRAS.328..726S, 2009MNRAS.399..497D}. For TNG100, the subhalo merger trees are constructed using the SubLink algorithm \citep{2015MNRAS.449...49R} whereas in EAGLE-L100 the merger trees are identified with the D-Trees algorithm \citep{Jiang_2014}.

\begin{figure}
    \centering
    \subfloat{\includegraphics[width=\linewidth]{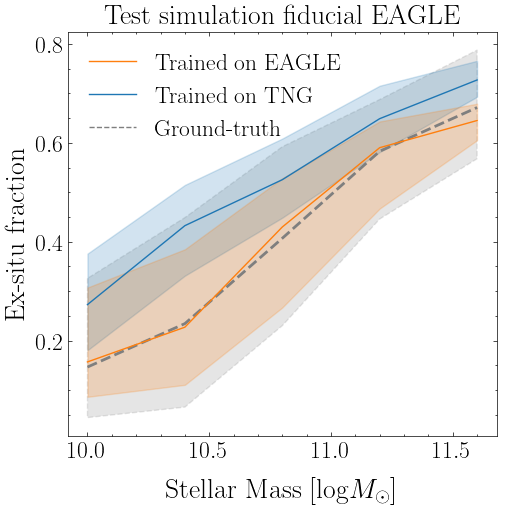}} 
    \caption{The recovery of the ex-situ stellar mass fraction vs. stellar mass relation from the CNN models trained on the TNG100 and EAGLE-L100 public release data when applied on the public EAGLE-L100 test set. Similar to Figure \ref{fig:mass_recovery}, we plot the recovery of the stellar mass vs. ex-situ fraction relation from models trained on mass, kinematics, age and metallicity 2D spatial maps within 4 $r_{e}$ aperture. Additional to the ground-truth, we plot the recovery from CNN models trained on the EAGLE-L100 with an orange line (median) and shaded region (68 per cent containment) as well as from CNN models trained on the TNG100 dataset (blue curve and shades). We find that the different definitions between simulations further bias results, as TNG100 constantly over-predicts the ex-situ stellar mass fraction of galaxies from the EAGLE-L100 test set by around 10-25 per cent for the considered stellar mass range. The domain drift caused by the different underlying sub-grid recipes is further enhanced by the diverging merger tree definitions.}
    \label{fig:mass_recovery_eagle_fid}
\end{figure}

To investigate the effect of the different merger tree algorithms, we repeat the dataset creation for the fiducial EAGLE-L100 in an identical fashion as already described in this work. We re-train the CNN models with the new datasets and cross-test them across simulations. We now find that the domain drift is significantly higher when the stellar mass, kinematics, age and metallicity spatial maps in a 4 $r_{e}$ aperture are used for inferring the ex-situ stellar mass fraction. In Figure \ref{fig:mass_recovery_eagle_fid}, we demonstrate this domain drift by showing the apparent over-prediction exhibited when testing the fiducial EAGLE-L100 test set on a model trained on TNG100, which now reaches error values $\sim30\%$. This suggests that the varying definitions between simulations add an extra layer of bias that can further impede inference across simulations. This result further underlines that simulation-based inference should always be approached with caution prior to application on actual observational data. Notably, we find once again that stellar mass and kinematic spatial maps prove to be the most robust predictors across simulations.

\section{Summary and Conclusions} \label{sec:conclusion}
In this work, we have utilized machine learning techniques along with two cosmological simulations, TNG100 and EAGLE, to infer the ex-situ stellar mass fraction of galaxies from observable quantities, such as the stellar mass, kinematics, age and metallicity, both as integrated properties and spatially-resolved two-dimensional maps. The use of two distinct hydrodynamical simulations, instead of one, is twofold. First, we wish to investigate whether and how the sub-grid physics introduced by each cosmological model affect the relations between observable properties and the merging history of a galaxy. Our end goal is to develop a robust model that surpasses the aforementioned differences and can be applied on actual observational data.

We choose a sample of galaxies from TNG100 and EAGLE-L100 with stellar mass $> {10}^{10}\ {M}_{\odot }$ from redshifts $z = 0, z = 0.1$ and  $z = 0.2$. We initially compare the ex-situ stellar mass fraction prediction with two approaches, namely inferring with integrated galaxy properties as an input as well as using 2D normalized maps of stellar mass, velocity, velocity $\sigma$, age and metallicity. These two methods require different neural network architectures. We train a probabilistic multilayer perceptron for the integrated inputs and a probabilistic convolutional neural network for the 2D spatial maps approach. The advantage of the latter approach is that it enables us to increase our training sample by using multiple realizations per galaxy through different projections. The probabilistic architecture allows a quantification of the uncertainty of each prediction. 

In a fixed simulation scenario, we find that the ex-situ stellar mass fraction can be inferred accurately from both the integrated galaxy inputs as well as from the 2D observable spatially-resolved maps, that have been normalized individually to contain only gradient information. Notably though, we report a significant increase in accuracy upon using the 2D observable inputs, especially in higher ex-situ stellar mass fractions. We proceed by cross-testing the trained neural networks across simulations to evaluate if the models are robust to the divergent underlying sub-grid recipes and find an apparent decrease in the predictive power of the models. Focusing our work on the 2D spatial maps approach henceforward, we investigate the origin of this domain drift between the two simulations by experimenting with the different input channels and the aperture. We find that:

\begin{itemize}
    \item The ex-situ stellar mass fraction can be accurately predicted from spatially-resolved 2D maps containing only gradients in a fixed simulation scenario, even when only the inner parts of galaxies are considered as inputs.
    \item Stellar mass and kinematics (V,  $\rm\sigma$) spatial maps are the most robust predictors across cosmological simulations.
    \item The inner parts of galaxies are adequate for recovering the global ex-situ stellar mass fraction and surprisingly minimize the domain drift across simulations. This is a welcome result, as the ultimate objective of this work is the application of the trained models on IFU surveys typically covering a narrower FOV.
    \item Training only with mass and kinematics spatial maps within 1 $r_{e}$ aperture provides unbiased predictions across simulations with <15 per cent scatter independent of the origin of the training inputs.
    \item The robustness of the models is verified by their ability to recover the ex-situ stellar mass fraction vs. stellar mass relation for both TNG100 and EAGLE-L100 irrespective of the training dataset.
\end{itemize}

Aiming to better understand the source of domain drift upon utilizing the two simulation datasets in their whole extent, we inspect the representation space the networks learn as well as the spatial gradients present in the input maps and find significant differences. The existence of populations with a particular merging history that are mapped to diverging observable properties is exhibited especially when stellar population properties are included in the inputs of the trained models. A closer inspection of the relation between the ex-situ stellar mass fraction and the gradients between different radii in the 2D spatial maps reveals that the metallicity gradients are inherently different among the two simulations, especially in the outer galactic regions. We conclude that, while metallicity provides important information for the inference of the ex-situ stellar mass fraction in a single simulation scenario, it is not a robust predictor across simulations. 

This is the first work that attempts to cross test neural networks across two different cosmological simulations to infer an aspect of the merging history of galaxies using machine learning. We have shown that the ex-situ stellar mass fraction can be predicted from observable 2D spatial maps with very high accuracy irrespective of the simulation dataset used for training when only mass and kinematic gradients within 1 $r_{e}$ aperture are used as input. We find that the achieved precision is comparable to the ones reported from previous works in a single simulation setup and surpassing by far the bare relation between ex-situ vs. stellar mass extracted directly from the simulations on focus. More importantly, this precision is reported from models trained on datasets stripped from all mass and size dependencies, underlining the power of gradients in resolving the merging history of galaxies.


In the near future, we plan to assess the robustness of the trained models to observational effects by utilizing fully forward modeled mock survey samples like MaNGIA \citep{mangia}.



\section*{Acknowledgements}

We thank the referee for their constructive comments which improved the clarity of this paper. JFB and EA acknowledge support through the RAVET project by the grant PID2019-107427GB-C32 from the Spanish Ministry of Science, Innovation and Universities (MCIU), and through the IAC project TRACES which is partially supported through the state budget and the regional budget of the Consejer\'ia de Econom\'ia, Industria, Comercio y Conocimiento of the Canary Islands Autonomous Community. MHC, RS and EA acknowledge financial support from the State Research Agency (AEI\-MCINN) of the Spanish Ministry of Science and Innovation under the grants ``Galaxy Evolution with Artificial Intelligence" with reference PGC2018-100852-A-I00 and "BASALT" with reference PID2021-126838NB-I00. This research made use of computing time available on the high-performance computing systems at the Instituto de Astrofisica de Canarias. The author thankfully acknowledges the technical expertise and assistance provided by the Spanish Supercomputing Network (Red Espanola de Supercomputacion), as well as the computer resources used: the Deimos-Diva Supercomputer, located at the Instituto de Astrofisica de Canarias.

\section*{Data Availability}

 Data directly related to this publication and its figures will be made available on request from the corresponding author. The outputs of the TNG (https://www.tng-project.org/) and EAGLE
(http://icc.dur.ac.uk/Eagle/) simulations are all publicly available. The associated code used to produce the results presented in this work is available at \url{https://github.com/eagel27/galaxy-exsitu}.



\bibliographystyle{mnras}
\bibliography{example} 




\appendix
\section{Description of model architectures and inputs}

We present two different approaches for inferring the ex-situ stellar mass fraction that require distinct neural network architectures. The architecture details of the MLP and the CNN models can be found in Table \ref{tab:neural_network_dense} and Table \ref{tab:neural_network_cnn} respectively. A short description of the integrated properties used for training the MLP models is provided in table \ref{tab:integral_inputs_table}. Correspondingly, information on which database fields were used for the creation of the 2D maps utilized as inputs of the CNNs can be found in Table \ref{tab:2d_maps_table}.

\begin{table}[ht]
    \centering
    \small
    \begin{tabular}{ p{3cm}|p{2cm}|p{2cm} }
        
        \textbf{Layer} & \textbf{Output Shape} & \textbf{Params} \\
        \hline
        Input & 5 & 0  \\\hline
        Dense & 32  & 288 \\\hline
        BatchNorm & 32  & 128 \\\hline
        Dropout & 32  & 0 \\\hline
        Dense & 16  & 528 \\\hline
        BatchNorm & 16  & 64 \\\hline
        Dropout & 16  & 0 \\\hline
        Dense & 16  & 272 \\\hline
        BatchNorm & 16  & 64 \\\hline
        Dropout & 16   & 0 \\\hline
        Flatten & 16 & 0 \\\hline
        Dense & 2 & 34 \\\hline
        IndependentNormal & & 0 \\
    \end{tabular}

    \caption{The Multilayer Perceptron architecture. The layer configuration of the MLP shows that the network receives the 5 integated values as an input and outputs a normal distribution.}
    \label{tab:neural_network_dense}
\end{table}

\begin{table}[ht]
    \centering
    \small
    \begin{tabular}{ p{3cm}|p{2cm}|p{2cm} }
        
        \textbf{Layer} & \textbf{Output Shape} & \textbf{Params} \\
        \hline
        Input & (128, 128, 5) & 0  \\\hline
        Conv2D & (128, 128, 32)   & 4032 \\\hline
        MaxPooling2D & (64, 64, 32)  & 0\\\hline
        BatchNorm & (64, 64, 32)  & 128 \\\hline
        Dropout & (64, 64, 32)  & 0 \\\hline
        Conv2D & (64, 64, 64)   & 18496 \\\hline
        MaxPooling2D & (32, 32, 64)  & 0\\\hline
        BatchNorm & (32, 32, 64)  & 256 \\\hline
        Dropout & (32, 32, 64)  & 0 \\\hline
        Conv2D & (32, 32, 64)  & 36928 \\\hline
        MaxPooling2D & (16, 16, 64)  & 0\\\hline
        BatchNorm & (16, 16, 64)  & 256 \\\hline
        Dropout & (16, 16, 64)   & 0 \\\hline
        Flatten & (16384, 1) & 0 \\\hline
        Dense & (512, 1) & 8389120 \\\hline
        BatchNorm & (512, 1) & 2048 \\\hline
        Dropout & (512, 1)   & 0 \\\hline
        Dense & (2, 1) & 1026 \\\hline
        IndependentNormal & & 0 \\
    \end{tabular}

    \caption{The CNN architecture utilized in this work. The model receives 5 input maps of size 128x128 and outputs a normal distribution after a succession of convolutional blocks.}
    \label{tab:neural_network_cnn}
\end{table}

\begin{table*}
    \centering
    \begin{tabular}{ p{1.8cm}p{6cm}p{3.5cm}p{3cm}  }
    \hline
    \multicolumn{4}{c}{Observable integrated properties} \\
    \hline
    Name & Description & DB field & Note\\
    \hline
    Stellar Mass & The total mass of all stellar particles & SubhaloMassType (Star)  & Subhalo catalog  \\

    Metallicity & Mass-weighted metallicity of each galaxy from the stellar particles
contained within two times the stellar half-mass radius. & GFM\_Metallicity  & Snapshot PartType4\\

    Age & Mass-weighted mean age of each galaxy from the stellar particles
contained within two times the stellar half-mass radius. & GFM\_StellarFormationTime & Snapshot PartType4\\

    Stellar half-mass radius & The radius containing half of the stellar mass of the galaxy (kpc)  & SubhaloHalfmassRadType (Star) & Subhalo catalog\\
        
    Spin &  Total spin per axis, computed for each as the mass weighted sum of the relative coordinate times relative velocity of all member particles/cells.  (kpc/h)(km/s)  & SubhaloSpin & Subhalo catalog \\
    \end{tabular}
    \caption{The galaxy observable integrated values used for inferring the ex-situ stellar mass fraction for the TNG100 and EAGLE-L100 simulations. For each property, a short description is given as well as the catalog field or snapshot field utilized to calculate the said property for both simulations.}
    \label{tab:integral_inputs_table}
\end{table*}

\begin{table*}
    \centering
    \begin{tabular}{ p{3cm}p{3.5cm}p{7cm}p{1.5cm}  }
    \hline
    \multicolumn{4}{c}{Observable properties} \\
    \hline
    2D Map & Stellar Particle property & Description & Binning \\
    \hline
    Mass & Masses &  Mass of this star or wind phase cell.  &Sum  \\
    Velocity & Velocities   & Spatial velocity. Multiply this value by $\sqrt{\alpha}$ to obtain peculiar velocity.  & Mean \\
    Velocity Dispersion & Velocities & Spatial velocity. Multiply this value by $\sqrt{\alpha}$ to obtain peculiar velocity.  & STD \\
    Metallicity & GFM\_Metallicity & The ratio $M_Z/M_{total}$ where $M_Z$ is the total mass all metal elements (above He). We convert to solar metallicity by dividing by 0.0127 (the primordial solar metallicity). &Mean \\
    Age &  GFM\_StellarFormationTime  & The exact time (given as the scalefactor) when this star was formed. We calculate the age of each stellar particle in Gyr from the scale factor in the $\Lambda$CDM cosmology context. &  Mean\\
    \end{tabular}
    \caption{The galaxy observable 2D maps used for inferring the ex-situ stellar mass fraction for the TNG100 and EAGLE-L100 simulations. For each resolved map, the stellar particle properties used for the creation are shown along with the operation required and a short description.}
    \label{tab:2d_maps_table}
\end{table*}

\bsp	
\label{lastpage}
\end{document}